\documentclass[final,times,fleqn,11pt]{elsarticle}
\biboptions{sort&compress}
\usepackage{amssymb,amsthm,mhchem}
\usepackage{dsfont}
\usepackage{booktabs,tabularx,graphicx}
\usepackage{bm}
\usepackage{color,soul,lineno,bm}
\usepackage[linesnumbered,ruled,vlined]{algorithm2e}
\usepackage[colorlinks=true, linkcolor=blue, urlcolor=blue, citecolor=blue]{hyperref}

\begin{document}

\begin{frontmatter}

\title{Phase-field analysis for brittle fracture in ferroelectric materials with flexoelectric effect}

\author[inst1]{Chang Liu\corref{cor1}}
\cortext[cor1]{Corresponding author. Email address: lc@swjtu.edu.cn}
\author[inst2]{Yu Tan}
\author[inst3]{Yong Zhang}
\author[inst1]{Zhaoyi Liu}
\author[inst4]{Takahiro Shimada}
\author[inst1]{Xiangyu Li}
\author[inst5]{Jie Wang}

\affiliation[inst1]{organization={School of Mechanics and Aerospace Engineering, Southwest Jiaotong University},
            city={Chengdu},
            postcode={610031}, 
            country={China}}
\affiliation[inst2]{organization={College of Environment and Civil Engineering, Chengdu
University of Technology},
            city={Chengdu},
            postcode={610059}, 
            country={China}}
\affiliation[inst3]{department={School of Aerospace Engineering and Applied Mechanics},
    organization={Tongji University},
            city={Shanghai},
            postcode={200092}, 
            country={China}}
\affiliation[inst4]{department={Department of Mechanical Engineering and Science},
organization={Kyoto University},
city={Kyoto},
postcode={615-8540},
country={Japan}}
\affiliation[inst5]{organization={Department of Engineering Mechanics, School of Aeronautics and Astronautics, Zhejiang University},
            city={Hangzhou},
            postcode={310027}, 
            country={China}}

\begin{abstract}
Understanding the nature of brittle failure in ferroelectric materials is essential, but difficult due to the complex interaction between mechanical and electrical concentrated fields near the crack tip. In this work, an extended phase-field model incorporating multiple order parameters is constructed to analyze the coupled evolution of fracture and domain behavior in ferroelectric materials. The strain gradient is incorporated into the governing equations to evaluate the impact of the flexoelectric effect during the crack propagation process. Our advanced phase-field model demonstrated that, with the consideration of the flexoelectric effect, both the crack extension rate and crack path are related to the initial polarization direction. This phenomenon is  associated with the eigenstrain induced by the flexoelectric effect. This study provides in-depth insight into the fracture behavior of ferroelectric materials. The developed model framework can also be employed to investigate electromechanical coupling failures in more complex ferroelectric structures.
\end{abstract}


\end{frontmatter}


\section{Introduction}

Ferroelectric materials have extensive applications to advanced functional components, including capacitors,  non-volatile memories and ultrasonic transducers \cite{martin_thin-film_2017}. Currently, most of the widely used ferroelectric materials like \ce{Pb[Zr_{(1-x)} Ti_x]O3} (PZT) ceramics and \ce{Pb(Mg1/3 Nb2/3)O3-PbTiO3} (PMN-PT) crystals are intrinsically brittle, with the strength generally less than 100 MPa, and the fracture toughness merely around {1 $\mathrm{MPa}\cdot\mathrm{m}^{1/2}$} \cite{schneider_influence_2007}. However, as the core of actuators and ultrasonic generators,  ferroelectric materials are often subjected to mechanical vibrations and high-frequency electric fields. Such a severe environment can easily induce the initiation and propagation of cracks inside the ferroelectric materials. Accordingly, a comprehensive study on the fracture behaviors of ferroelectric materials is necessary for the reliable design of ferroelectric devices. 

The initiation and propagation of crack in ferroelectric material is accompanied by the microstructure evolution of polarization domain morphologies \cite{yingwei_influence_2020}. Such an electromechanical coupling property makes the fracture behaviors of ferroelectric materials rather complex. Thus, numerous experimental tests have been carried out for a better understanding of the fracture mechanism. For example, the fracture toughness of poled ferroelectric materials exhibits a significant anisotropic property in the Vickers indentation test \cite{yamamoto_internal_1983,okazaki_electro-mechanical_1992,lynch_crack_1995}. With the utilization of a high-speed camera, Wang et al. \cite{wang_influence_2020} captured the process of crack propagation in PZT-5H samples during the three-point bending fracture experiment. 

Strain gradient exist around the crack tip due to the formation of a singular stress field, leading to a significant flexoelectric effect \cite{zubko_flexoelectric_2013,yudin_fundamentals_2013}. Since ferroelectric materials with higher dielectric constants have large flexoelectric coefficients \cite{ma_flexoelectricity_2006,hong_first-principles_2013}, the crack tip is subjected to a giant strain gradient (flexoelectric) field, leading to a complex evolution of domain structure during the crack propagation process. For example, Wang et al. \cite{wang_direct_2020} observed a large polarization response near the crack tip in \ce{SrTiO3} epitaxial thin film induced by flexoelectric effect. Recently, Edwards et al. \cite{cordero-edwards_flexoelectric_2019} found that the flexoelectric effect around the crack tip can induce anisotropy in the fracture behavior of ferroelectric materials, and the crack propagation path shows asymmetrical characteristics.  Using an atomic force microscopy, Xu et al. \cite{xu_directly_2023} determined the electric field distribution around the crack tip, which provides a reliable approach to measuring the flexoelectric effect in ferroelectric materials. 

To gain a thorough understanding of the microscopic fracture mechanism in ferroelectric materials, the in situ observation and capturing of electromechanical response is indispensable. Currently, these requirements are still challenging for experimental investigation. Hence, numerical approach can further enrich the understanding of the fracture behavior in ferroelectric materials as complements to the experimental method. In recent years, the phase-field model has gained much attention in the study of microstructure evolution. The phase-field model utilizes a diffuse interface instead of a sharp one, which eliminates the requirements to track the moving of interface. Thus, the computation is greatly simplified \cite{chen_phase-field_2002}. A phase-field model has been widely used to investigate the polarization evolution in ferroelectrics \cite{wang_phase-field_2004,su_continuum_2007,hong_stability_2017,yuan_defect-mediated_2018}. Wang et al. \cite{wang_phase_2007,wang_phase_2008,wang_three-dimensional_2009,wang_effect_2010} utilized the phase-field to investigate the impact of domain switching on fracture toughness with different initial polarization directions, electric fields, and electrical boundary conditions. Using the phase-field model, Yu \cite{yu_i-integral_2016} et al. developed an I-integral method to analyze the crack-tip intensity factor, which is independent of domain switching and is suitable for large-scale domain switching. Zhao et al. \cite{zhao_effect_2018} incorporated the flexoelectric effect into the phase-field model for ferroelectric materials, which suggests that the flexoelectric effect can induce an asymmetric distribution of domain structures near the crack tip.

With the combination of Griffith fracture theory, the phase-field model can also be used to capture the fracture behaviors in ferroelectric materials. In the context of the linear piezoelectric theory, Tan et al. \cite{tan_phase_2023,tan_phase_2022} employed the phase-field model to investigate the fracture process of multiferroic materials  and the fatigue fracture behavior of ferroelectric materials. Taking the polarization evolution into account, Xu et al. \cite{xu_fracture_2010} established a phase-field continuum model for ferroelectric materials with a damage variable. Abdollahi et al. \cite{abdollahi_phase-field_2011,abdollahi_phase-field_2011-1,abdollahi_phase-field_2012,abdollahi_three-dimensional_2014} conducted a series of phase-field models to study the coupling between crack propagation and domain evolution in ferroelectric single crystals and polycrystals. Zhang et al. \cite{yong_zhang_phase_2022,zhang_jumping_2022} introducing the dielectric breakdown as an additional order parameter in the phase-field model. The evolution of polarization, dielectric breakdown, and crack propagation can consequently be calculated simultaneously. Nevertheless, the flexoelectric effect has yet to be incorporated into the phase-field model for the fracture of ferroelectric materials to the best of the author's knowledge.

In this work, a phase-field model for the brittle fracture in ferroelectric materials with the consideration of the flexoelectric effect is developed. To satisfy the high-order continuity requirements of spatial discretization, the isogeometric analysis method is employed. Additionally, polynomial splines over hierarchical T-meshes (PHT-splines) are used to enable local mesh refinement near the crack, thereby enhancing computational efficiency. With this model, the crack evolution with a different initial polarization direction is simulated. The effect of polarization domain switching and flexoelectricity throughout the crack extension process is investigated. 

The structure of this paper is organized as follows. Section \ref{Section2} introduces the framework of the phase-field model and the corresponding governing equations. The numerical implementation is detailed in Section \ref{Section3}. Section \ref{Section4} presents the simulation results and discussions. The concluding remarks are draw in Section\ref{Section5}.
\section{Formulation}\label{Section2}
To investigate the fracture behaviors in ferroelectrics, it is critical to capture both the evolution of the domain and crack. Consequently, polarization and fracture are utilized as order parameters to describe the coupling evolution.  In this section, the two phase-field models for ferroelectrics and brittle fracture are introduced, respectively. Subsequently, their natural coupling and numerical implementation are discussed.

\subsection{Phase-field model for ferroelectric materials}
For a ferroelectric material free of body charge and body force, the electric enthalpy density is formulated by \cite{gu_flexoelectricity_2014,jiang_polarization_2015}:
\begin{equation}\label{eq1}
    \psi(\mathbf{\varepsilon},\nabla\mathbf{\varepsilon},\phi,\mathbf{P},\nabla\mathbf{P})
    =\psi_\mathrm{Land}+\psi_\mathrm{grad}+\psi_\mathrm{elas}+\psi_\mathrm{flex}+\psi_\mathrm{elec},
\end{equation}
where $\mathbf{\varepsilon}=\frac{1}{2}\left[\nabla \mathbf{u}+(\nabla\mathbf{u})^\mathrm{T}\right]$ is the total strain, $\nabla=\frac{\partial ()}{\partial x_j}\mathbf{e}_j $ is the gradient operator, $\mathbf{u}$, $\phi$ and $\mathbf{P}$ represent the displacement vector, electric potential, and polarization vector, respectively.

The first term on the right side of Eq.\eqref{eq1} is the Landau energy density:
\begin{equation}
   \begin{split}
     \psi_\mathrm{Land}&=\bar{\mathbf{\alpha}}:\mathbf{P}\otimes\mathbf{P}+\bar{\bar{\mathbf{\alpha}^*_4}} \mathbf{P}\otimes\mathbf{P}\otimes\mathbf{P}\otimes\mathbf{P}+\bar{\bar{\bar{\mathbf{\alpha}^*_6}}}\mathbf{P}\otimes\mathbf{P}\otimes\mathbf{P}\otimes\mathbf{P}\otimes\mathbf{P}\otimes\mathbf{P}\\
    &+\bar{\bar{\bar{\bar{\mathbf{\alpha}^*_8}}}} \mathbf{P}\otimes\mathbf{P}\otimes\mathbf{P}\otimes\mathbf{P}\otimes\mathbf{P}\otimes\mathbf{P}\otimes\mathbf{P}\otimes\mathbf{P},
   \end{split}
\end{equation}
where $\bar{\mathbf{\alpha}}$ is the dielectric stiffness linearly depending upon temperature, $\bar{\bar{\mathbf{\alpha}}}$, $\bar{\bar{\bar{\mathbf{\alpha}}}}$ and $\bar{\bar{\bar{\bar{\mathbf{\alpha}}}}}$ are the forth, sixth, and eighth order dielectric stiffness coefficients, respectively. The symbol $\otimes$ and  $\overset{*}{n}$ are dyadic operation and multiple dot product, respectively. For ferroelectric materials, expanding the polynomial to the $8^\mathrm{th}$-order can accurately describe the ferroelectric phases with monoclinic and higher symmetries \cite{heitmann_thermodynamics_2014}.

In the context of the Ginzburg-Landau theory, the electric enthalpy density also depends upon the gradient of polarization. In ferroelectrics, the gradient energy density $\psi_\mathrm{grad}$ represents the energy stored in the polarization domain walls, which can be expressed as:
\begin{equation}
    \psi_\mathrm{grad}=\frac{1}{2}\nabla\mathbf{P}:\mathbb{G}:\nabla \mathbf{P},
\end{equation}
where $\mathbb{G}$ is the $4^\mathrm{th}$-rank gradient energy tensor. 

The elastic energy density $\psi_\mathrm{elas}$ reads
\begin{equation}\label{f_elas}
    \psi_\mathrm{elas}=\frac{1}{2}\mathbf{\varepsilon}_\mathrm{e}:\mathbb{C}:\mathbf{\varepsilon}_\mathrm{e},
\end{equation}
with $\mathbb{C}$ donating the elastic stiffness tensor. The elastic strain $\mathbf{\varepsilon}_\mathrm{e}=\mathbf{\varepsilon}-\mathbf{\varepsilon}_0$, where $\mathbf{\varepsilon}_0$ is spontaneous strain of ferroelectric material related to the spontaneous polarization:
\begin{equation}\label{eq5}
    \mathbf{\varepsilon}_0=\mathbb{Q}:\mathbf{P}\otimes\mathbf{P},
\end{equation}
with $\mathbb{Q}$ denoting the electrostrictive coefficient tensor.

The flexoelectric energy $\psi_\mathrm{flex}$ can be written as 
\begin{equation}
    \psi_\mathrm{flex}=\frac{1}{2}(\nabla \mathbf{P}: \mathbb{F}: \mathbf{\varepsilon}-\mathbf{P}\cdot\mathbb{F}\vdots\nabla\mathbf{\varepsilon}),
\end{equation}
where $\mathbb{F}$ is the flexoelectric tensor. 

The last term in Eq.\eqref{eq1} is the electric energy:
\begin{equation}
    \psi_\mathrm{elec}=-\frac{1}{2}\mathbf{E}\cdot \mathbf{K} \cdot \mathbf{E}-\mathbf{E}\cdot\mathbf{P},
\end{equation} 
where $\mathbf{K}$ represents the dielectric coefficient, and $\mathbf{E}=-\nabla \phi$ is the electric field with $\phi$ denoting the electric potential.

\subsection{Phase-field model for brittle fracture}
Currently, the commonly used ferroelectric materials are ceramics and crystals, which are susceptible to brittle fracture. The phase-field model for brittle fracture is adopted in this paper, which is derived from Griffith fracture theory \cite{francfort_revisiting_1998, bourdin_numerical_2000}. To approximate the discontinuous crack surface, a phase-field variable $\nu$ is introduced. This variable facilitates the representation of the crack as a damage band with a finite width. The variable $\nu$ continuously changes from 0 to 1, representing the process of the material from being intact to completely fractured. 

In the phase-field model for brittle fracture, the total enthalpy density of the structure should include the fracture energy density,
\begin{equation}
    \psi_{\mathrm{t}}=\psi_0+\psi_{\mathrm{frac}},
\end{equation}
in which $\psi_0$ is the bulk enthalpy density of ferroelectric material with crack, and $\psi_\mathrm{frac}$ represents the fracture energy related to the energy dissipation in creating the crack surface:
\begin{equation}\label{eq9}
    \psi_{\mathrm{frac}}(v,\nabla v)=G_c\left[\frac{1}{2l_c}\nu^2+\frac{l_c}{2}\nabla v\cdot \nabla v\right].
\end{equation}
In this context,  $G_c$ refers to the critical energy release rate, which is equivalent to twice the surface energy according to the Griffith theory. The width of the damage band is controlled by the positive regularization parameter $l_c$. The width of the damage band reduces to 0 as $l_c$ approaches 0, and the model also approximates the sharp fracture theory \cite{miehe_thermodynamically_2010}.

By introducing the fracture energy, the phase-field model transforms the issue of crack initiation and propagation  into an optimization problem of seeking the minimum energy under multi-field coupling conditions. Therefore, it can be used to directly solve complex fracture problems without additional crack path tracking methods.

\subsection{Brittle fracture in ferroelectrics}
The phase-field model for brittle fracture turns out to be a continuum damage model. In regions with non-zero $\nu$, the material exhibits varying degrees of damage. This damage affects the electromechanical behavior of ferroelectric material and dissipates part of the electric enthalpy. Furthermore, the evolution of cracks creates new interfaces within the material. Accurately describing the boundary conditions of these internal crack surfaces is crucial to developing a phase-field fracture model. To investigate the crack evolution in ferroelectrics, it is necessary to determine the coupling relationship among the crack and other physical fields, such as strain $\mathbf{\varepsilon}$, electric field $\mathbf{E}$, and polarization $\mathbf{P}$. 

In this study, a degradation function $g(\nu)$ is adopted to characterize the dissipation of enthalpy resulting from damage in ferroelectric materials \cite{abdollahi_phase-field_2011,xu_fracture_2010}. Generally, the degradation function should satisfy the following requirements \cite{miehe_thermodynamically_2010}:
\begin{enumerate}
    \item $g(0) = 1 $: This implies no energy dissipation in undamaged regions.
    \item $g(1) = 0 $: This denotes complete material damage, characterized by the loss of electromechanical properties.
    \item $g(1)' = 0$: This condition signifies regions of total fracture, where no further evolution of the phase-field occurs.
    \item $g(\nu)'<0$ if $\nu\neq 1$: This indicates that as damage increases, the degradation function diminishes, rendering $g(\nu)$ a strictly monotonically decreasing function with respect to $\nu$.
\end{enumerate}
In view of these considerations, this study utilizes a degradation function in the form of $g(\nu)=(1-\nu)^2+\eta$, where $\eta$ is a small quantity to avoid singularities in the calculation. 

Integrating the degradation function into the elastic energy density yields:
\begin{equation}
    \psi_{\mathrm{elas}}^\mathrm{d}=g(\nu)\psi_{\mathrm{elas}}.
\end{equation}
This approach ensures that the mechanically traction-free boundary conditions on crack surfaces, denoted as $\mathbf{\sigma}\cdot\mathbf{n}=0$, are satisfied within the framework of phase-field model. Here, $\mathbf{\sigma}$ and $\mathbf{n}$ represent the mechanical stress and the unit outward normal vector, respectively.

Due to the symmetry breaking at the crack surface, the polarization is influenced by the surface effect. In this study, In this study, the free-polarization boundary condition is applied on the crack surface:
\begin{equation}
    \frac{\mathrm{d}P_i^+}{\mathrm{d}\mathbf{n}^+}=\frac{\mathrm{d}P_i^-}{\mathrm{d}\mathrm{\mathbf{n}^-}}=0,
\end{equation}
where the superscripts "+" and "-" indicate the upper and lower surfaces of the crack, respectively. In the phase-field model, the free-polarization boundary condition is incorporated by multiplying the gradient energy with the degradation function $g(\nu)$, as shown in:
\begin{equation}
    \psi_\mathrm{grad}^\mathrm{d}=g(\nu)\psi_\mathrm{grad}.
\end{equation}
Given that the flexoelectric energy includes both strain and polarization gradients, it is similarly adjusted by multiplying by the degradation function:
\begin{equation}
    \psi_{\mathrm{flex}}^\mathrm{d}=g(\nu) \psi_{\mathrm{flex}}.
\end{equation}
Conversely, the Landau energy, which does not contain polarization gradient terms, remains unaltered, as noted in Ref. \cite{abdollahi_phase-field_2011}.

The electric boundary condition is crucial to determine the fracture behavior of ferroelectric materials. In the fracture models for ferroelectrics, there are two classical extreme assumptions, namely electrically permeable and impermeable boundary conditions. Under the permeable boundary condition, the electric potential and electric displacement are identical on both sides of the crack \cite{parton_fracture_1976}:
\begin{equation}
    \phi^+=\phi^-,\; \mathbf{D}^+\cdot\mathbf{n}^+=\mathbf{D}^-\cdot\mathbf{n}^-,
\end{equation}
The electric energy $\psi_{\mathrm{elec}}$ remains unchanged for the continuity of electric fields. On the other hand, the electrically impermeable boundary condition implies a charge-free crack surface:
\begin{equation}
    \mathbf{D}^+\cdot\mathbf{n}^+=\mathbf{D}^-\cdot\mathbf{n}^-=0.
\end{equation}

In comparison to the electrically permeable crack, there is an absence of electric displacement within the cracked region in the electrically impermeable crack1. Consequently, the electric energy should be modified by  incorporating the degradation function, as shown in the equation \cite{wang_effect_2010}:
\begin{equation}
    \psi_{\mathrm{elec}}^\mathrm{d}=g(\nu)\psi_{\mathrm{elec}}.
\end{equation}
In the case of an impermeable crack, which represents an open and electrically defective crack, the electric field becomes discontinues across the crack surface.

In short, for ferroelectric materials, the total enthalpy density for the electrically permeable crack and the electrically impermeable crack can be written respectively as:
\begin{subequations}
    \begin{equation}
        \psi_\mathrm{t}=g(\nu)\left( \psi_\mathrm{grad}+\psi_\mathrm{elas}+\psi_\mathrm{flex} \right)+\psi_\mathrm{Land}+\psi_\mathrm{elec}+\psi_\mathrm{frac},
    \end{equation}
    \begin{equation}
        \psi_\mathrm{t}=g(\nu)\left( \psi_\mathrm{grad}+\psi_\mathrm{elas}+\psi_\mathrm{flex}+\psi_\mathrm{elec} \right)+\psi_\mathrm{Land}+\psi_\mathrm{frac}.
    \end{equation}
\end{subequations}
Here, the coupling between the fracture field variable $\nu$ and other physical fields $\mathbf{\varepsilon}$, $\mathbf{\phi}$, and $\mathbf{P}$ have been constructed. 

\subsection{Governing equations}
The constitutive equations of each physical field can be derived from the total enthalpy density $\psi_\mathrm{t}$:
\begin{equation}
    \mathbf{\sigma}=\frac{\partial \psi_\mathrm{t}}{\partial\mathbf{\varepsilon}},\; 
    \mathbf{\tau}=\frac{\partial \psi_\mathrm{t}}{\partial \nabla\mathbf{\varepsilon}},\; 
    \mathbf{\eta}=\frac{\partial\psi_\mathrm{t}}{\partial \mathbf{P}},\;
    \mathbf{\Lambda}=\frac{\partial \psi_\mathrm{t}}{\partial \nabla\mathbf{P}},\;
    \mathbf{D}=\frac{\partial \psi_\mathrm{t}}{\partial \mathbf{E}},
\end{equation}
where $\mathbf{\sigma}$, $\mathbf{\tau}$, $\mathbf{\eta}$, $ \mathbf{\Lambda}$, and $\mathbf{D}$ denote the mechanical stress tensor, the higher order stress, the effective local electric force, the higher order local electric force and the electric displacement, respectively. The thermodynamic consistency of the constitutive equations incorporating the flexoelectric effect has been proven by Gu et al. \cite{gu_flexoelectricity_2014} and Tagantsev et al. \cite{tagantsev_flexoelectric_2012}.

In absence of the body force and body charge, the stress and electric displacement filed of a ferroelectric material should satisfy the continuum mechanical equilibrium equations and the Maxwell equation \cite{chang_liu_isogeometric_2019}:
\begin{equation}\label{equilibrium}
    \nabla\cdot\left( \mathbf{\sigma}-\nabla\cdot\mathbf{\tau} \right)=0\; \mathrm{in}\; \Omega,\;  
\end{equation}
\begin{equation}\label{Maxwell}
    \nabla \cdot \mathbf{D}=0 \; \mathrm{in}\; \Omega,\;
\end{equation}
with the Dirichlet and Neumann-type boundary conditions:
\begin{equation}
    \begin{split}
         (\mathbf{\sigma}-\nabla\cdot\mathbf{\tau})\cdot \mathbf{n}=\mathbf{t} \;\mathrm{on}\; \partial\Omega_t,\; 
        & \mathbf{\tau}:\mathbf{n}\otimes \mathbf{n} = \mathbf{M} \; \mathrm{on}\; \partial \Omega_M,
        &\mathbf{u}=\mathbf{u}^*\; \mathrm{on}\; \partial\Omega_u, \;\\
         \mathbf{D}\cdot \mathbf{n}=-\omega\;\mathrm{on}\; \partial\Omega_\omega,\;
        &\phi=\phi^* \;\mathrm{on}\; \partial\Omega_\phi,
    \end{split}
\end{equation}
where $\partial\Omega_t$, $\partial\Omega_M$, $\partial\Omega_u$, $\partial\Omega_\omega$, and $\partial\Omega_\phi$ are boundaries of the ferroelectric body $\Omega$ with applied traction, higher order traction, displacement, free charge and electric potential, respectively.

Ferroelectric materials typically form a domain structure spontaneously below the Curie temperature, characterized by polarization arranged in a head-to-tail configuration in most circumstances. When subjected to external stress fields or electric fields, the domain morphology will evolve with the motion and nucleation of domain walls. The temporal and spatial evolution of polarization domain is governed by the time-depenedent Ginzbureg Landau equation:
\begin{equation}\label{TDGL}
    \dot{\mathbf{P}}=-L_P\left[ \frac{\partial \psi_t}{\partial \mathbf{P}}-\nabla \cdot \left( \frac{\partial\psi_t}{\partial \nabla\mathbf{P}} \right) \right],
\end{equation}
where $\dot{\mathbf{P}}=\partial \mathbf{P}/\partial t$ represents the time derivative of polarization, and $L_P$ is the kinetic coefficient for polarization evolution. 

The crack propagation process is described in Ref.\cite{abdollahi_phase-field_2011}:
\begin{equation}\label{phase-field crack1}
    \dot{\nu}=-L_\nu \left[ \frac{\partial\psi_t}{\partial v}-\nabla \cdot \left( \frac{\partial \psi_t}{\partial \nabla \nu} \right) \right].
\end{equation}
In Eq.\eqref{phase-field crack1}, the term $\partial\psi_t/\partial v$ describes the influence of displacement, electric field, and polarization on the fracture evolution. Since crack propagation is inherently coupled with domain evolution, a finite kinetic coefficient $L_\nu$ is incorporated to capture the dynamic switching of polarization during crack growth.

Selecting an appropriate driving force is essential for accurately predicting fractures. Experimental research by Park and Sun \cite{park_fracture_1995} demonstrated that the fracture behavior in piezoelectric material is primarily driven by mechanical forces. Inspired by this finding, only the elastic energy is employed to drive the crack propagation in the phase-field model \cite{wu_phase-field_2021}. Therefore, Eq.\eqref{phase-field crack1} should be adjusted as:
\begin{equation}\label{phase-field crack2}
    \dot{\nu}=L_\nu\left[ 2(1-\nu) \psi_\mathrm{elas}+G_c\left(\frac{\nu}{l_c} +l_c\Delta \nu\right) \right].
\end{equation}

While the driving force for crack propagation is described in Eq.\eqref{phase-field crack2}, it does not distinguish fracture behaviors under tensile and compressive stresses. Miehe et al. \cite{miehe_thermodynamically_2010} introduced a strain energy decomposition method utilizing the spectral decomposition of the strain tensor. In this method, the elastic strain tensor $\mathbf{\varepsilon}_\mathrm{e}$ is decomposed into the positive and negative components as:
\begin{equation}
    \mathbf{\varepsilon}_\mathrm{e}=\mathbf{\varepsilon}^+_\mathrm{e} + \mathbf{\varepsilon}^-_\mathrm{e}  ,
\end{equation}
where 
\begin{equation}
    \mathbf{\varepsilon}_\mathrm{e}^\pm=\sum_{d}^{a=1}\langle \varepsilon_a\rangle^\pm \mathbf{n}_a\otimes\mathbf{n}_a,
\end{equation}
with $\varepsilon_a$ and $\mathbf{n}_a$ representing the eigenvalues and eigenvectors of the strain $\mathbf{\varepsilon}$, respectively. For simplicity, the Macaulay brackets $\langle x \rangle^\pm$ are introduced: 
\begin{equation}
    \langle x \rangle^\pm = \frac{1}{2}(x\pm |x|).
\end{equation}
Then, the elastic energy can be decomposed as:
\begin{equation}
    \psi_\mathrm{eals}=\psi_\mathrm{eals}^++\psi_\mathrm{eals}^-,
\end{equation}
with 
\begin{subequations}
    \begin{equation}
        \psi^+_\mathrm{elas}=\frac{1}{2}\mathbf{\varepsilon}_\mathrm{e}^+ : \mathbb{C} : \mathbf{\varepsilon}_\mathrm{e}^+,
    \end{equation}
    \begin{equation}
        \psi^-_\mathrm{elas}=\frac{1}{2}\mathbf{\varepsilon}_\mathrm{e}^- : \mathbb{C} : \mathbf{\varepsilon}_\mathrm{e}^-.
    \end{equation}
\end{subequations}
In the decomposed strain energy density, only the positive part $\psi^+_\mathrm{elas}$ drives the crack propagation, while the negative part $\psi^-_\mathrm{elas}$ does not induce changes in the crack. 

To prevent crack healing during the unloading process, $\nu$ should maintain monotonic increase. Therefore, a historical variable $H$ is induced:
\begin{equation}
    H(\mathbf{x},t)=\max_{t\in [0,t_n]} \left\{  \psi^+_\mathrm{elas}(\mathbf{x})\right\},
\end{equation}
which represents the maximum value of $\psi^+_\mathrm{elas}$ over the time interval $[0,t_n]$. Consequently, Eq.\eqref{phase-field crack2} should be modified:
\begin{equation}
    \dot{\nu}=L_\nu\left[ 2(1-\nu)H+G_c\left(\frac{\nu}{l_c} + l_c\Delta \nu\right) \right].
\end{equation}

An additional constraint is employed in case of the inter penetration of crack face in compression:
\begin{equation}
     \nu(x)=0, \forall x: \psi^+(x)<\psi^-(x).
\end{equation}

For some models, it is necessary to set the initial crack. Within the framework of the phase-field method, three typical approaches are available \cite{sargado_high-accuracy_2018}: 
\begin{enumerate}
    \item Geometrically defined crack: Initial cracks are represented as discrete cracks within the model's geometry.
    \item Phase-field defined crack: Initial cracks are given by the Dirichlet boundary conditions, where $\nu=1$ in the cracked region and $\phi=0$ in all other regions. 
    \item Historically defined crack: Initial cracks are modeled by the local historical variable $H(\mathbf{x},0)$
    \begin{equation}\label{historical variable}
        H(\mathbf{x},0)=B_c\left\{ 
            \begin{aligned}
                &\frac{G_c}{2l_0}(1-\frac{2d(\mathbf{x},l)}{l_0}) & d(\mathbf{x},l)\leq \frac{l_0}{2} \\
                &0 & d(\mathbf{x},l) > \frac{l_0}{2}
            \end{aligned}
         \right.,
    \end{equation}
    where $d(\mathbf{x},l)$ is the distance from point $\mathbf{x}$ to the initial crack $l$.
\end{enumerate}

Modeling the initial crack through the phase-field method tends to accelerate crack growth, whereas other approaches may slow down the process. In this work, the historically defined crack is used with the scalar parameter $B=1000$ for all the simulations \cite{borden_phase-field_2012}.
\section{Numerical implementation}\label{Section3}
\subsection{Isogeometric analysis}
With the presence of the flexoelectric effect, the strain gradients appear in the governing equations Eq.\eqref{equilibrium}-\eqref{phase-field crack2}. Thus, the globally $C^1$-continuous basis function is required to guarantee the accuracy of spatial discretization. However, constructing such functions within the classical finite element method is still a significant challenge. As an alternative approach, the isogeometric analysis method (IGA) is employed to numerically solve the governing equations in this work. The polynomial splines over hierarchical T-meshes (PHT-spline) are adopted as the basis function to achieve high-order continuity \cite{deng_polynomial_2008}. comparison to the commonly used non-uniform rational B-spline (NURBs) in IGA, the PHT-spline is more convenient for implementing local mesh refinement, enhancing the computational efficiency for complex problems \cite{anitescu_recovery-based_2018}. Moreover, PHT-splines facilitate automatic mesh refinement around the crack propagation path, which can balance solution accuracy and computational efficiency \cite{goswami_adaptive_2020}.

PHT-splines extend the concept of B-splines to T-meshes. The knot vectors $U^i$ are defined as 
\begin{equation}
    U^i=\{\xi_1^i, \xi_2^i, \ldots, \xi_{n+1}^i\} \quad i\in\{1,2,3\},
\end{equation}
where $0=\xi_1^i\leq\xi_2^i\leq \ldots \leq\xi_{n+1}^i=1$ for each parameter direction $i$, and $n_i$ denotes the number of elements along each parameter direction.

For the two-dimensional problems, the mesh $\mathbb{T}_0$ in the parameter space is constructed as
\begin{equation}
    \mathbb{T}_0=\{E_{0,k_m}=[\xi_{k_1-1}^{(1)},\xi_{k_1}^{(1)}]\times[\xi_{k_2-1}^{(2)},\xi_{k_2}^{(2)}], k_1=2,\ldots,n_1+1,\;  k_2=2,\dots,n_2+1\},
\end{equation}
where $E_{0,k_m}$ represents the $k_m$-th element in the mesh at level 0 with $k_m=(k_2-2)n_1+(k_1-1)$. The Bézier expression of the basis function $N_{l,k_m}$ is given by
\begin{equation}
    N_{l,k_m}(\xi^{(1)},\xi^{(2)})=\sum_{i_1=1}^{p+1}\sum_{i_2=1}^{p+1}C^E_{i_1 i_2,k_m}\hat{B}_{i_1 i_2}\circ\hat{\mathbf{F}}^{-1}(\xi^{(1)},\xi^{(2)}),
\end{equation}
in which $C^E_{i_1 i_2,k_m}$ is the Bézier coefficients of the basis functions, $\hat{B}_{i_1 i_2}$ denotes the tensor product of Bernstein polynomial, and $\hat{\mathbf{F}}$ represents the linear mapping from the reference interval $[-1,1]$ to the element $E_{l, k_m}$. The Bernstein polynomial in one dimension, $B_{i,p}(\xi)$, is defined as:
\begin{equation}
    B_{i,p}(\xi)=2^{-p}\left( \begin{matrix}
        p \\
        i-1
    \end{matrix} \right)(1-\xi)^{p-i+1}(1+\xi)^{i-1},\; i=1,2,\ldots,p+1,
\end{equation}
with $p$ indicating the order of the Bernstein polynomial. For a comprehensive exploration of PHT-spline and local mesh refinement strategies, please refer to Ref. \cite{goswami_adaptive_2020,anitescu_recovery-based_2018}.

\subsection{Discretization}
In the IGA implementation, the displacement $\mathbf{u}$, the electric potential $\phi$, the polarization $\mathbf{P}$, and the fracture order parameter $v$ are taken as the nodal degree of freedom, which can be discretized as 
\begin{equation}\label{discretizedfields}
    \mathbf{u}=\sum_{i=1}^n \mathbf{N}_i\mathbf{u}_i,\;
    \phi=\sum_{i=1}^n N_i\phi_i,\;
    \mathbf{P}=\sum_{i=1}^n \mathbf{N}_i\mathbf{P}_i,\;
     \nu=\sum_{i=1}^n \mathbf{N}_i\nu_i,
\end{equation}
where $\mathbf{u}_i$, $\phi_i$, $\mathbf{P}_i$, and $\nu_i$ represent the field variables at the $i$-th control point, and $N_i$ is the corresponding basis function, $n$ is the total number of basis functions associated with the element. For simplicity, the two-dimensional model is taken as an example. The matrix $N_i$ is defined as 
\begin{equation}
    \mathbf{N}_i=\left[ \begin{matrix}
        N_i & 0 \\
        0 & N_i
    \end{matrix} \right].
\end{equation}
The strains, electric fields, polarization gradients, fracture order parameter gradients, and strain gradient can be approximated by
\begin{equation}
\begin{split}\label{eq40}
    & \mathbf{\varepsilon}=\sum^n_{i=1}\mathbf{B}^u_i u_i,\;
        \mathbf{E}=-\sum^n_{i=1}\mathbf{B}^\phi_i \phi_i, \;
        \nabla\mathbf{P}=\sum^n_{i=1}\mathbf{B}^P_i\mathbf{P}_i,\\
    & \nabla \nu= \sum^n_{i=1} \mathbf{B}^{\nu}_i\nu_i,\;
        \nabla\mathbf{\varepsilon}=\sum^n_{i=1} \mathbf{H}_i \mathbf{u}_i,\;
\end{split}
\end{equation}
where the gradient operators $\mathbf{B}^u_i$, $\mathbf{B}^\phi_i$, $\mathbf{B}^P_i$, $\mathbf{B}^\nu_i$, and the Hessian operator $\mathbf{H}_i $ can be expressed as following
\begingroup
\renewcommand{\arraystretch}{1.5}
\begin{equation}\label{HB}
    \begin{split}
       & \mathbf{B}^u_i=\left[ \begin{matrix}
            \frac{\partial}{\partial x} & 0 \\
            0 &  \frac{\partial}{\partial y} \\
            \frac{\partial}{\partial y} &  \frac{\partial}{\partial x}
        \end{matrix} \right]\mathbf{N}_i, \;
        \mathbf{B}^\phi_i=\left[\begin{matrix}
            \frac{\partial}{\partial x} \\
            \frac{\partial}{\partial y}
        \end{matrix}\right]N_i,\\
        &\mathbf{B}^P_i=\left[ \begin{matrix}
            \frac{\partial}{\partial x} & 0 \\
            0 &  \frac{\partial}{\partial y} \\
            \frac{\partial}{\partial y} & 0 \\
            0 &  \frac{\partial}{\partial x} \\
        \end{matrix} \right]\mathbf{N}_i, \;
        \mathbf{B}^\nu_i=\left[\begin{matrix}
            \frac{\partial}{\partial x} \\
            \frac{\partial}{\partial y}
        \end{matrix}\right]N_i, \\
        &\mathbf{H}_i=\left[ \begin{matrix}
            \frac{\partial^2}{\partial x^2} & \frac{\partial^2}{\partial x \partial y} & 0 & 0 & \frac{\partial^2}{\partial x \partial y} & \frac{\partial^2}{\partial y^2} \\
            0 & 0 & \frac{\partial^2}{\partial x \partial y} & \frac{\partial^2}{\partial y^2} & \frac{\partial^2}{\partial x^2} & \frac{\partial^2}{\partial x \partial y}
        \end{matrix} \right]^T\mathbf{N}_i.
    \end{split}
\end{equation}
\endgroup

According to the variation principle of virtual work, the weak form of governing equation can be obtained as 
\begin{subequations}\label{weakform}
    \begin{equation}\label{weak1}
        \int_\Omega \left( \mathbf{\sigma}:\nabla\delta\mathbf{u}+\mathbf{\tau}\vdots\nabla\delta\mathbf{\varepsilon} \right)\mathrm{d}V-\int_{\partial \Omega_t}\mathbf{t}\cdot \delta \mathbf{u}\,\mathrm{d}S - \int_{\partial\Omega_M}\mathbf{M}:\delta\mathbf{\varepsilon}\,\mathrm{d}S=0,
    \end{equation}
    \begin{equation}\label{weak2}
        \int_\Omega\mathbf{D}\cdot\nabla\delta\phi\,\mathrm{d}V+\int_{\partial\Omega_\omega}\omega\delta\phi\,\mathrm{d}S=0,
    \end{equation}
    \begin{equation}\label{weak3}
        \int_\Omega \left[ \left(\mathbf{\eta}+\frac{\dot{\mathbf{P}}}{L_P}\right)\cdot \delta \mathbf{P}
        +\mathbf{\Lambda}:\nabla\delta\mathbf{P} \right] \mathrm{d}V=0,
    \end{equation}
    \begin{equation}\label{weak4}
        \int_\Omega\left[ 2(\nu-1)\left( H+\frac{G_c}{l_c} \right)+\frac{\dot{\nu}}{L_\nu}\right]\delta\nu\,\mathrm{d}V+\int_\Omega L_\nu G_c l_c \nabla \nu \cdot \nabla \delta \nu \,\mathrm{d}V=0.
    \end{equation}
\end{subequations}

Substituting Eqs.\eqref{discretizedfields}-\eqref{HB} into Eqs.\eqref{weakform}, the weak form of the governing equations is reformulated into matrix notation. From these expressions, both the residual and the tangent matrices can be derived, which are listed in the supplementary material.

To align with existing phase-field models for ferroelectric materials \cite{chang_liu_isogeometric_2019}, this study adopts a staggered algorithm to solve Eqs.\eqref{weakform}. The process starts by addressing the displacement, electric potential, and polarization fields through Eqs.\eqref{weak1}-\eqref{weak3}, which employs the backward Euler method for temporal integration. The nonlinear equations at each iteration are solved by the Newton-Raphson method. Upon achieving a steady state, the computed values for $\mathbf{u}$, $\phi$, and $\mathbf{P}$ are  utilized in Eq.\eqref{weak4} to determine $\nu$ at the current timestep. The details of the staggered algorithm are outlined in Algorithm \ref{flowchart}.

\begin{algorithm}[!htbp]
    \SetKwData{Left}{left}\SetKwData{This}{this}\SetKwData{Up}{up}
    \SetKwFunction{Union}{Union}\SetKwFunction{FindCompress}{FindCompress}
    \SetKwInOut{Input}{Input}\SetKwInOut{Output}{Output}
    \Input{$\mathbf{u}_{n-1}$, $\phi_{n-1}$, $\mathbf{P}_{n-1}$, $\nu_{n-1}$}
    \Output{$\mathbf{u}_{n}$, $\phi_{n}$, $\mathbf{P}_{n}$, $\nu_{n}$}
    \BlankLine
    Set boundary conditions $\mathbf{t}$, $\mathbf{M}$, $\mathbf{u}^*$, $\omega$, and $\phi^*$ \;
    Set tolerance $\delta_P$ and $\delta_\nu$ \;
    $\mathbf{u}_{n}^0=\mathbf{u}_{n-1}$, $\phi_{n}^0=\phi_{n-1}$, $\mathbf{P}_{n}^0=\mathbf{P}_{n-1}$, $\nu_{n}^0=\nu_{n-1}$\;
    \Repeat{$\|\mathbf{P}_{n}^{k+1}-\mathbf{P}_{n}^{k}\|_\infty<\delta_P$ and  $\|\nu_{n}^{k+1}-\nu_{n}^{k}\|_\infty<\delta_\nu$ }{Calculate $\mathbf{u}^{k+1}_{n}$, $\phi^{k+1}_{n}$, and $\mathbf{P}^{k+1}_{n}$ from Eqs.\eqref{weak1}-\eqref{weak3} with $\nu_{n}^k$\;
    Update the historical variable $H$\;
    Calcluate $\nu_{n}^{k+1}$ from Eq.\eqref{weak4} with $\mathbf{u}^{k+1}_{n}$, $\phi^{k+1}_{n}$, and $\mathbf{P}^{k+1}_{n}$\; }
    $\mathbf{u}_{n}=\mathbf{u}_{n}^{k+1}$, $\phi_{n}=\phi_{n}^{k+1}$, $\mathbf{P}_{n}=\mathbf{P}_{n}^{k+1}$, $\nu_{n}=\nu_{n}^{k+1}$

    \caption{Flowchart of the staggered scheme}\label{flowchart}
\end{algorithm}
\section{Results and discussions}\label{Section4}
\begin{figure}[htbp!]
    \centering
    \includegraphics[width=0.8\linewidth]{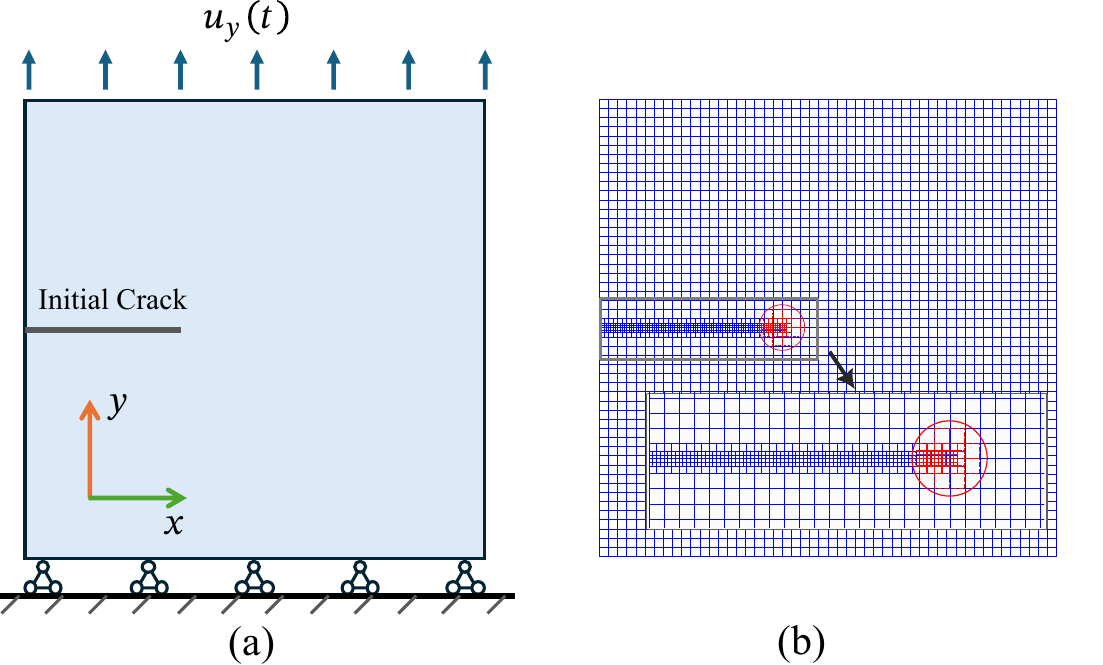}
    \caption{(a) Two-dimensional rectangular plate model and its corresponding boundary conditions. The dimension of the plate is 100 nm $\times$ 100 nm with a 40 nm pre-existing crack initiating from the left edge. Displacement along the $x_2$-axis is fixed at the bottom surface, while a time-dependent displacement, $u_2 (t)$, is imposed on the top surface. The origin of the coordinate system is located in the bottom left corner of the model. (b)  Mesh detail for the model. The discretized mesh is $50\times50$, with local refinement in the pre-existing crack area. The area enclosed by the gray box is an enlarged view of the local mesh around the crack. The regions marked by red circles and elements denote the area employed for calculating the configurational force.}
    \label{fig1}
\end{figure}

For simplicity, the present work adopts a two-dimensional plane stress model. A rectangular plate with a pre-existing edge crack serves as an example, with its dimensions and boundary conditions shown in Figure \ref{fig1}(a). The pre-existing crack is defined by the local historical variable $H$, which is expressed in Eq.\eqref{historical variable}. Dimensionless parameters are utilized to ensure computational accuracy \cite{chang_liu_isogeometric_2019}. The top edge of the plate is subjected to a time-dependent displacement load, $u_y^*=0. 002\times t^*$ is applied for the quasi-static model, where $t^*$ represents time, and the asterisk (*) denotes a dimensionless quantity. The simulation of an applied external electric field is achieved by establishing varying electric potentials across specific boundaries.  Unless specifically stated, it is assumed that both mechanical and electrical boundary conditions are traction-free and electrically open-circuited. The model uses a discretization of $50 \times 50$ elements, with local refinement at the crack region, as illustrated in Figure \ref{fig1}(b).  The material chosen for analysis is single-crystal \ce{PbTiO3} with its parameters listed in Table \ref{table1}. 
 
This section will primarily investigate two types of models: 1. The crack propagation direction is parallel to the polarization direction. This includes cases where both the crack and polarization directions align along the $x_1$-axis, and the case where the crack direction is opposite to the polarization direction (with polarization opposing the $x_1$-axis). 2. The crack propagation direction is perpendicular to the polarization direction, including cases with the polarization along the $x_2$-axis and cases with  polarization opposing the $x_2$-axs. In each case, domain evolution is initiated through a Gaussian random distribution of polarization. Before fracture evolution, polarization experiences enough time to relax sufficiently and form a stable domain structure. An impermeable crack boundary is employed to mimic air-filled cracks \cite{wang_effect_2010}. 

\begin{table}[!htbp]
    \centering
    \caption{Values of the dimensionless coefficients for \ce{PbTiO3}\cite{yong_zhang_phase_2022,eliseev_defect-driven_2018}}
    \label{table1}
    \begin{tabular}{lclcll}
        \toprule
        \multicolumn{6}{c}{Landau coeff.}                                                   \\ \midrule
        $\alpha_1$ & $\alpha_{11}$ & $\alpha_{12}$ & $\alpha_{111}$   & $\alpha_{112}$  &   \\
        -1.0          & -0.24            & 2.53             & 0.49               & 1.2               &   \\
        \multicolumn{3}{c}{gradient coeff.}        & \multicolumn{2}{c}{kinetic coeff.} &   \\ \cmidrule(r){1-3}\cmidrule(l){4-5}
        $G_{11}$   & $G_{12}$      & $G_{44}$      & $L_P$            & $L_\nu$         &   \\
        1.6          & 0             & 0.8             & 1.0                & 0.001               &   \\
        \multicolumn{2}{c}{elastic coeff.}        & \multicolumn{2}{c}{electrostrictive coeff.}      & \multicolumn{2}{c}{flexoelectric coeff.} \\ \cmidrule(r){1-2}\cmidrule{3-4}\cmidrule(l){5-6}
        $c_{11}$   & 1766             & $Q_{11}$      & 0.05                & $f_{11}$        & 51.7 \\
        $c_{12}$   & 802              & $Q_{12}$      &  -0.015              & $f_{12}$        & 37.3 \\
        $c_{44}$   & 1124             & $Q_{44}$      &  0.038              & $f_{44}$        & 42.6 \\
        \multicolumn{2}{c}{relative permittivity} & \multicolumn{2}{c}{critical energy release rate} & \multicolumn{2}{c}{length parameter}      \\ \cmidrule(r){1-2}\cmidrule{3-4}\cmidrule(l){5-6}
        $\kappa$   & 66            & $G_c$         & 20.2               & $l_0$           & 1 \\ \bottomrule
        \end{tabular}
\end{table}

\subsection{Crack parallel to the initial polarization direction}
To establish a single domain state with polarization vectors aligned along the $x_1$-axis, an external electric field $E_{1}^*=1$ is applied. This is achieved by setting the electric potential at the left and right edges to 0 and 100, respectively. 

\begin{figure}[!htbp]
    \centering
    \includegraphics[width=\linewidth]{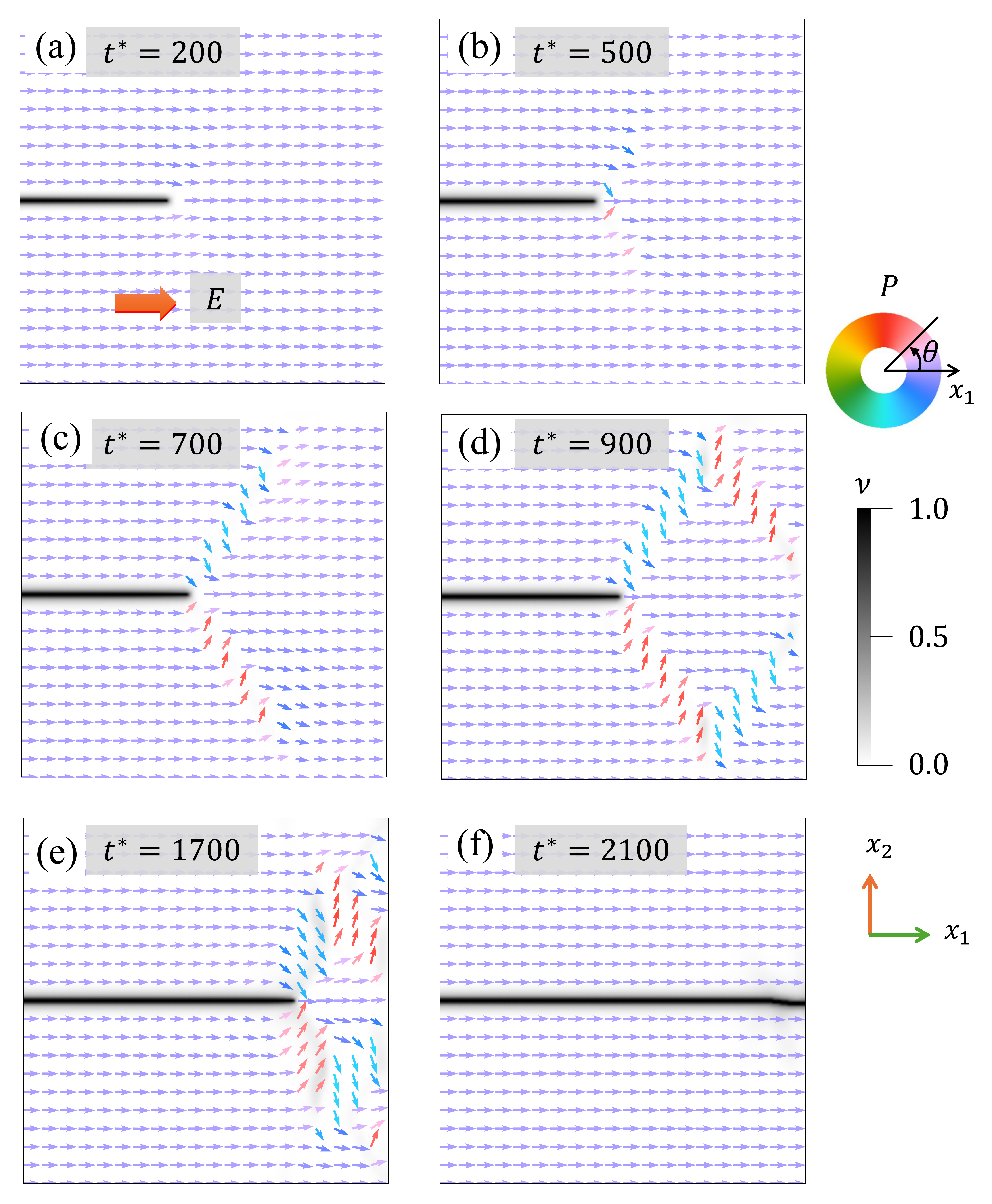}
    \caption{Dynamic progressions of crack propagation and polarization switching across various time intervals. The grayscale maps illustrate the spatial distribution of the fracture order parameter $\nu$. The arrows indicate the orientation and magnitude of the polarization vectors, with the colors representing the angle of polarization relative to the $x_1$-axis.} \label{fig2}
\end{figure} 

Figure \ref{fig2} presents the temporal progression of the domain structure alongside crack propagation without flexoelectric effect ($f_{ij}=0$). Before crack propagation, the material maintains almost a single domain state. In the vicinity of the crack tip, some spontaneous polarization rotate by angles less than $45^\circ$, yet no distinct switched zone forms, as illustrated in Figure \ref{fig2}(a). As the load $u_2^*$ increases, the pre-existing crack begins to extend. A distinctly switched zone develops in front of the crack tip, which is distinguished by two forward wings. The shape of the switched zone is almost symmetric with respect to the crack, while the directions of polarization switching within it are notably asymmetric. Specifically, the polarization in the wing above the crack rotates $90^\circ$ clockwise, whereas those below the crack rotate $90^\circ$ counterclockwise. The magnitude of $u_2^*$ directly influences both the width and length of the switched zone, with a larger value of $u_2^*$ leading to a more expansive region, as demonstrated in Figures \ref{fig2}(b)-(c). Similar domain evolution process near the crack tip are also reported by Wang et al. \cite{wang_phase_2008}.  

As $u_2^*$ continues to increase, the crack further propagates, and the switched zone expands correspondingly. Once the foremost area of the switched zone reaches the edge of the plate, a reflection-like phenomenon occurs, causing the zone to expand forward at the same reflection angle, and the angle of polarization rotation changes symmetrically. In the newly formed switched zone above the crack, the polarization rotates $90^\circ$ clockwise, whereas the polarization below the crack rotates $90^\circ$ counterclockwise. At this stage, the overall domain structure within the plate still exhibits symmetry along the crack plane. Among the four switched zones, the polarization alternately aligns in the up and down directions, forming a diamond structure that encloses the unswitched region, as illustrated in Figure \ref{fig2}(d). With further increases in $u_2^*$, the envelope area of the switched zone gradually decreases, while the width of the switch zones slightly expands. Symmetrically, $90^\circ$ polarization switched zones also emerge in the upper right and lower right corners of the plate, as presented in Figure \ref{fig2}(e). Subsequently, as the load continues to increase, the crack propagates to the right boundary of the plate, resulting in a complete fracture. At this point, the two sections of the plate along the crack plane can be considered as two independent plates. The displacement load no longer affects the distribution of the domain structure, which is now entirely controlled by the electric field, leading to the formation of two single-domain regions oriented along the $x_1$-direction, as shown in Figure \ref{fig2}(f). Throughout the loading process, polarization switching maintains symmetry relative to the crack plane, and the crack propagation path remains a straight path. It is only at moments close to complete fracture that minor differences in the domain structures on either side of the crack amplify, leading to a deviation in the crack propagation path. The complete spatial and temporal evolution of the domain structure and crack path is presented in Video S1.

\begin{figure}[!htbp]
    \centering
    \includegraphics[width=\linewidth]{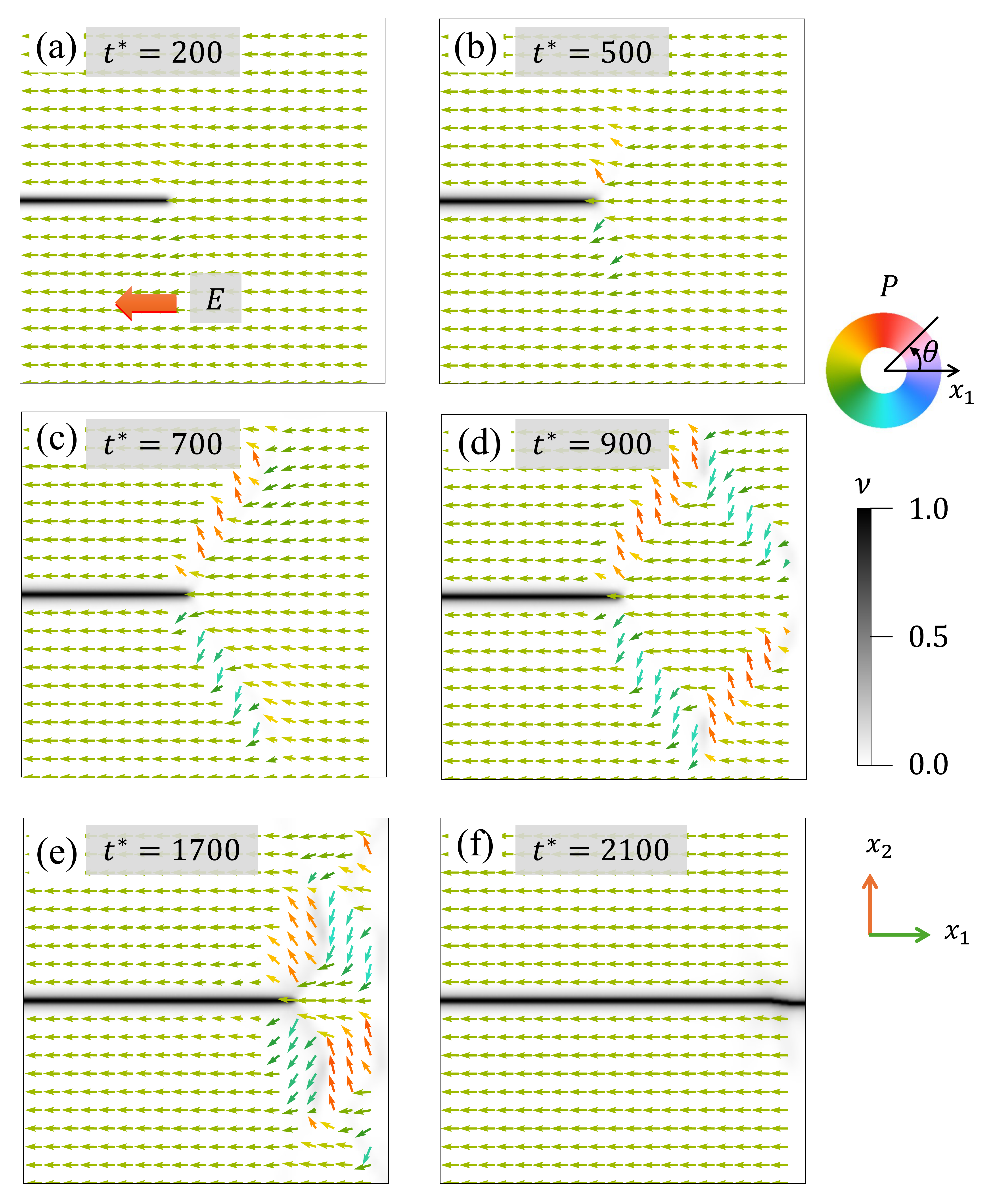}
    \caption{Morphologies of the crack and polarization at various moments $t^*=200$, 500, 700, 900, 1700, and 2100 under the applied electric field $E^*_1=-1$, respectively. Here, the flexoelectric effect is neglected.} \label{fig3}
\end{figure}

As a comparison, Figure \ref{fig3} illustrates the evolution of the crack and domain structure under an external load with the initial polarization opposing the $x_1$-axis. Similarly, the flexoelectric coefficients $f_{ij}$ are set to zero. The results at time steps $t^*=$200, 500, 700, 900, 1700, and 2100 show that both the length of crack propagation and the polarization evolution are identical with the case where polarization directions align along the $x_1$-axis. The difference lies in the polarization direction being opposite to the corresponding positions in Figure \ref{fig2}. The results indicate that, in the absence of the flexoelectric effect, as long as the polarization direction is parallel to the crack, the fracture process in ferroelectric materials remains consistent. The full spatial and temporal evolution of the domain structure and crack path are shown in Video S2.

\begin{figure}[!htbp]
    \centering
    \includegraphics[width=0.7\linewidth]{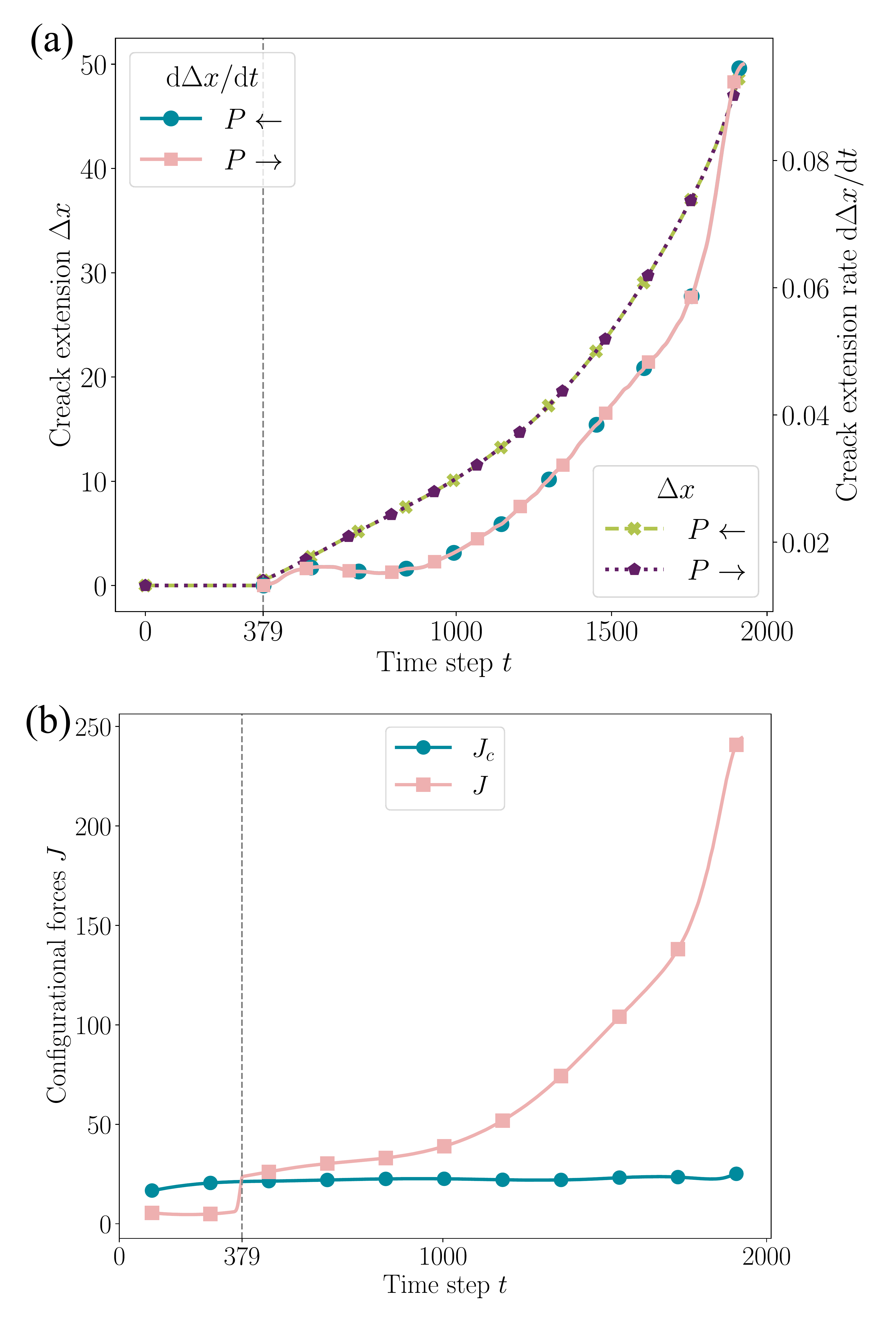}
    \caption{(a) The variation of crack extension $\Delta x$ (left $y$-axis) and the crack extension rate $\mathrm{d}\Delta x/\mathrm{d}t$ (right $y$-axis) with respect to the time step $t$. The arrows in the legend indicate the initial direction of polarization. Solid lines represent crack extension $\Delta x$, while dashed lines corresponding to the crack extension rate $\mathrm{d}\Delta x/\mathrm{d}t$; (b) The change of the crack extension driving force $J$ and local fracture toughness over time step $t$.} \label{fig4}
\end{figure}

Figure \ref{fig4}(a) illustrates the variations of crack extension length $\Delta x$ and the rate $\mathrm{d}\Delta x/\mathrm{d} t$ as functions of $t$ when the initial polarization is parallel to the crack. In the absence of the flexoelectric effect, the crack extension rate is identical in both cases where the initial polarization is oriented in the left or right direction. At $t^*=379$, which corresponds to the displacement load $u_y^*=0.794$, the crack begins to extend. From $t^*=400$ to $t^*=900$, the crack tip moves forward at a steady rate. Beyond the time step $t^*=900$, the crack extension rate increases rapidly, leading to an  accelerated crack propagation until complete fracture occurs.

To quantify the driving force for crack propagation in ferroelectric materials, the concept of configurational force is employed \cite{kuhn_discussion_2016}:
\begin{equation}
    \mathbf{g}_\mathrm{frac}=-\nabla\cdot\left( \psi_\mathrm{{frac}}\mathbf{I}-\nabla\nu\otimes\frac{\partial\psi}{\partial\nabla\nu} \right),
\end{equation}
\begin{equation}
    \mathbf{g}_{d}=-\frac{\dot{\nu}}{L_\nu}\nabla\nu,
\end{equation}
where the symbol $\mathbf{I}$ denote the 2nd-rank unit tensor. The terms $\mathbf{g}_\mathrm{frac}$ and $\mathbf{g}_d$ represent the resistance and net driving force for fracture, respectively. Aligning the concept of fracture configurational force with the $J$-integral, the summation of the $x_1$ components of $\mathbf{g}_\mathrm{frac}$ and  $\mathbf{g}_d$ can be represented by $J_c$ for the local fracture toughness, and by $J$ for the driving force, as detailed by Zhang et al.\cite{yong_zhang_phase_2022,zhang_jumping_2022}. The region for calculating $J$ and $J_c$ is selected by a circle centered at the crack tip, with a radius of 5, as illustrated in Figure \ref{fig1}(b).

Figure \ref{fig4}(b) shows the variation of the crack driving force $J^*$ and the resistance $J_c^*$ over time steps with $f_{ij}=0$. Throughout the entire crack propagation process, the resistance $J_c^*$ remains consistently stable at around 20, closely approximating the critical fracture toughness value of $G_c^*=20.2$.  Prior to the onset of crack propagation, the crack driving force $J^*$ is lower than the resistance $J_c^*$. As the applied displacement load $u_y^*$ continues to increase, $J^*$ exceeds $J_c^*$ at time $t^*=379$, triggering the start of crack growth. As the crack progresses from $t^*=400$ to $t^*=900$, new domain walls are established near the crack tip, as shown in Figure \ref{fig2} and Figure \ref{fig3}. During this period, the energy is allocated not only to the propagation of the crack but also to the generation of domain walls. This leads to a competition between $\psi_\mathrm{grad}$ and $\psi_\mathrm{frac}$, decelerating the crack extension rate, which is referred to as the polarization switching-induced toughening effect \cite{wang_phase_2007}. In the absence of new domain wall formation after $t^*>900$, the competition between $\psi_\mathrm{grad}$ and $\psi_\mathrm{frac}$ dissipates. Furthermore, the polarization switched zone continuously diminishes as the crack propagates, leading to a rapid increase in the $J^*$. This in turn causes the crack propagation rate to continually rise until complete fracture.

\begin{figure}[!htbp]
    \centering
    \includegraphics[width=\linewidth]{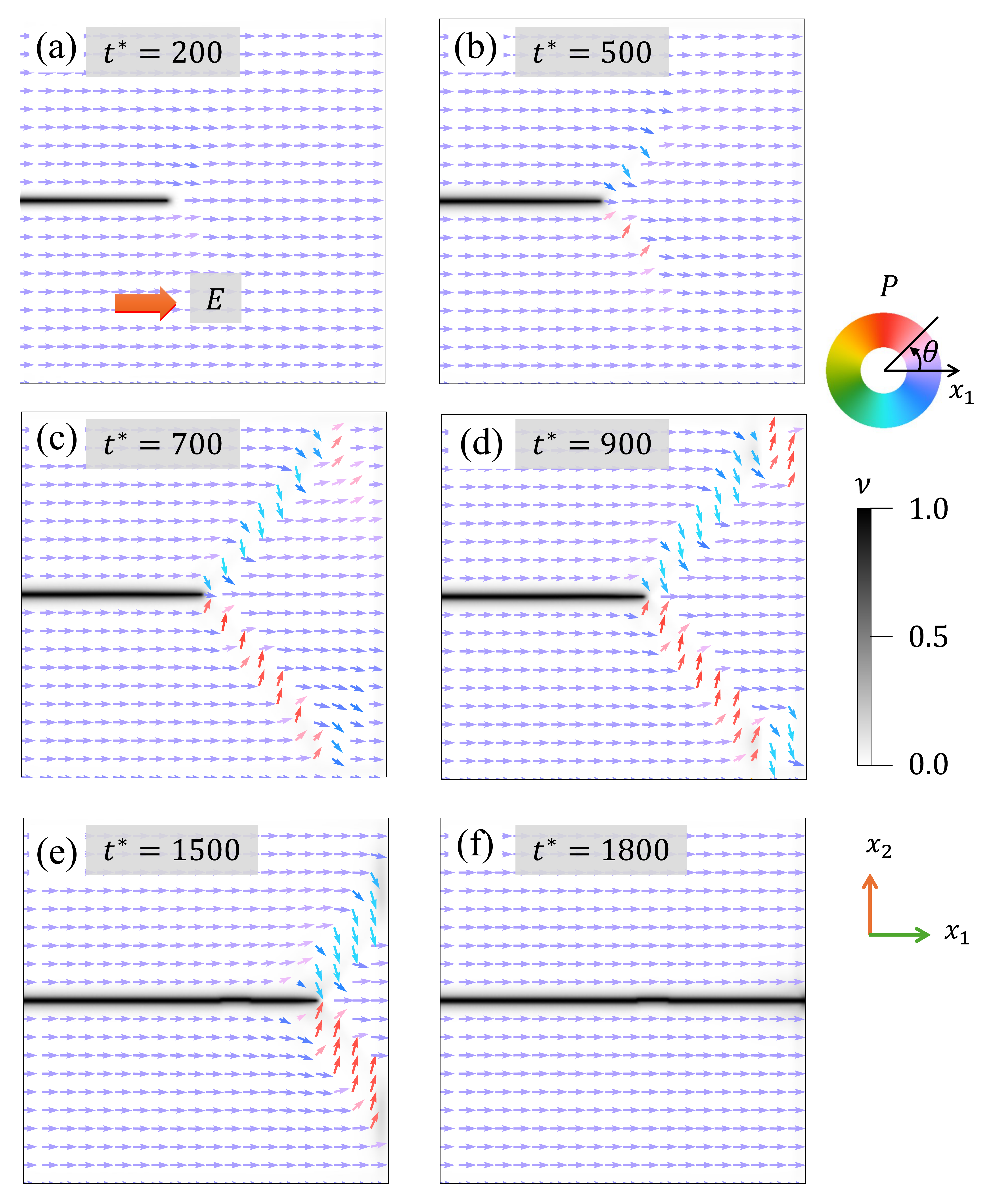}
    \caption{Morphologies of the crack and polarization at various moments $t^*=200$, 500, 700, 900, 1500, and 1800 under the applied electric field $E^*_1=1$, respectively. The flexoelectric effect is incorporated here.} \label{fig5}
\end{figure}

With the flexoelectric effect, the fracture behaviors in ferroelectric materials undergo noticeable changes. Figure \ref{fig5} presents the evolution of the crack path and domain structure in a ferroelectric plate across various time steps with the flexoelectric effect. The initial direction of polarization is oriented in the right direction. Similarly, before crack propagation, only the polarization near the crack tip undergoes rotations of less than $45^\circ$ without forming a distinct switched zone, as shown in Figure \ref{fig5}(a). Once the crack begins to extend, a switched zone with two forward wings also forms near the crack tip, as depicted in Figure \ref{fig5}(b) and (c). However, in contrast to the results in Figure 2, when the foremost area of the switched zone reaches the edge of the plate, the length of crack extension is significantly longer than the counterpart without the flexoelectric effect. Consequently, the front of the newly formed switched zone reaches the right edge of the plate, which is unable to form a complete diamond structure, as shown in Figure \ref{fig5}(d). As the crack continues to propagate, the two forward wings on both sides of the crack tip move forward, while the newly formed switched zone continually shrinks until it disappears, as illustrated in Figure \ref{fig5}(e). Once the plate is completely fractured, the polarization returns to a single domain state, as presented in Figure \ref{fig5}(f). The complete evolution process of the domain structure and crack path is presented in Video S3.

\begin{figure}[!htbp]
    \centering
    \includegraphics[width=\linewidth]{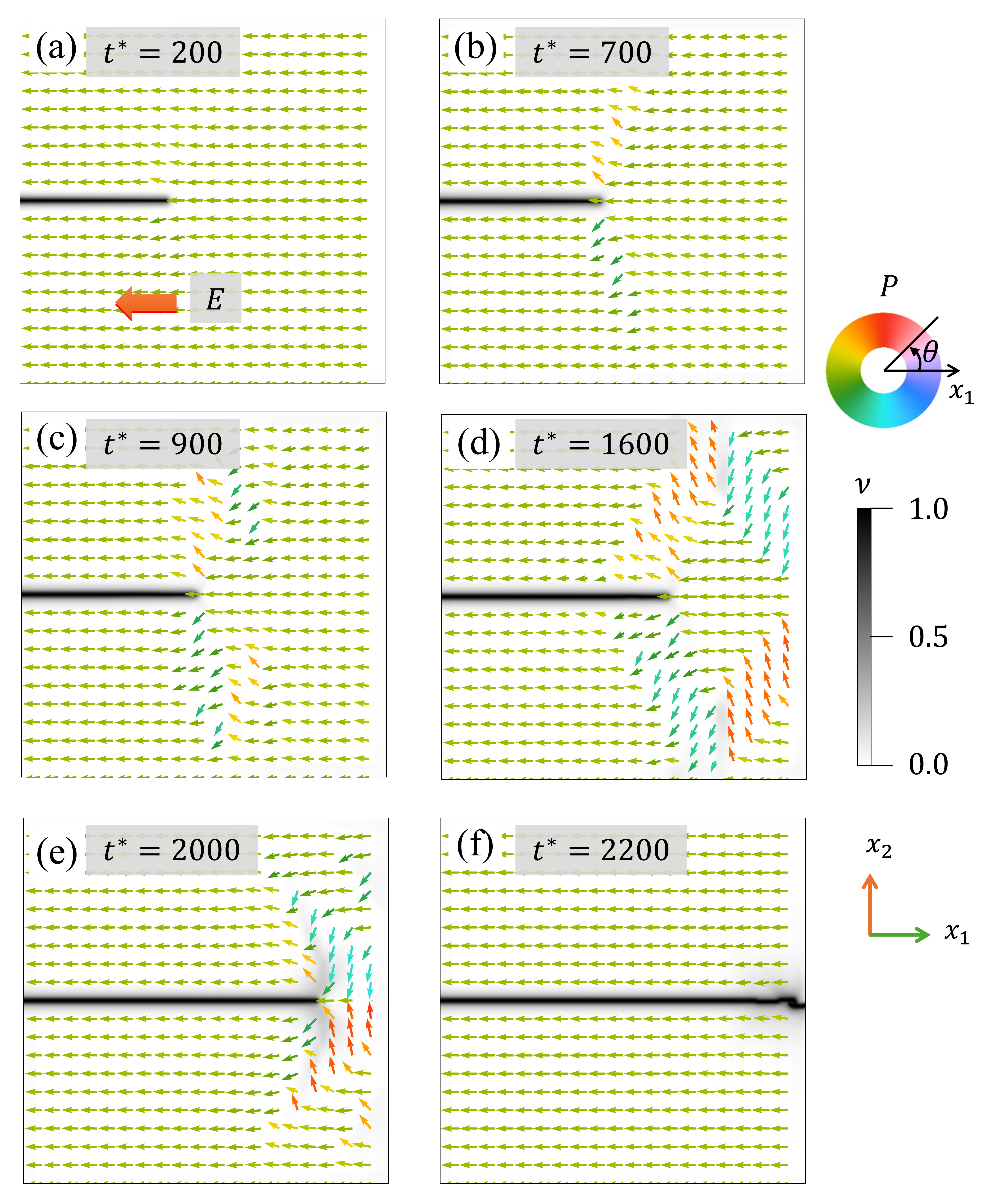}
    \caption{Morphologies of the crack and polarization at moments $t^*=200$, 700, 900, 1600, 2000, and 2200 under the applied electric field $E^*_1=-1$, respectively.} \label{fig6}
\end{figure}

Figure \ref{fig6} demonstrates the case where the initial polarization is opposite to the crack propagation with the consideration of the flexoelectric effect. As shown in Figure \ref{fig6}(a), prior to crack extension, the entire ferroelectric material maintains a single domain state with polarization aligned along the left direction, which is consistent with the direction of the electric field. With the crack extension, a $90^\circ$ rotation of polarization occurs on both sides of the crack tip, as depicted in Figure \ref{fig6}(b). This leads to the emergence of two symmetrical forward wings structures, with opposite polarization directions on each side. By the time $t^*=900$, the $90^\circ$ polarization switched zones develop on either side of the tips of the forward wings structure, leading to the formation of domain structures. Then compared to other cases, the ferroelectric material contains more domain walls, as displayed in Figure \ref{fig6}(c). At $t^*=1600$, the regions of polarization switching also evolve into diamond structures. However, unlike in other cases, the polarization switching zones started behind the crack tip rather than at the crack tip, as shown in \ref{fig6}(d). As the crack tip gradually approaches the right edge, the diamond structure vanishes, and the polarization switching region at the crack tip transforms into a backward wing structure. This structure maintains symmetrical in both morphology and polarization direction relative to the crack plane, with symmetrical auxiliary wing-like structures forming on both the upper and lower sides, as shown in Figure 6(e). The complete spatial and temporal evolution of the domain structure and crack path is presented in Video S4.

\begin{figure}[!htbp]
    \centering
    \includegraphics[width=\linewidth]{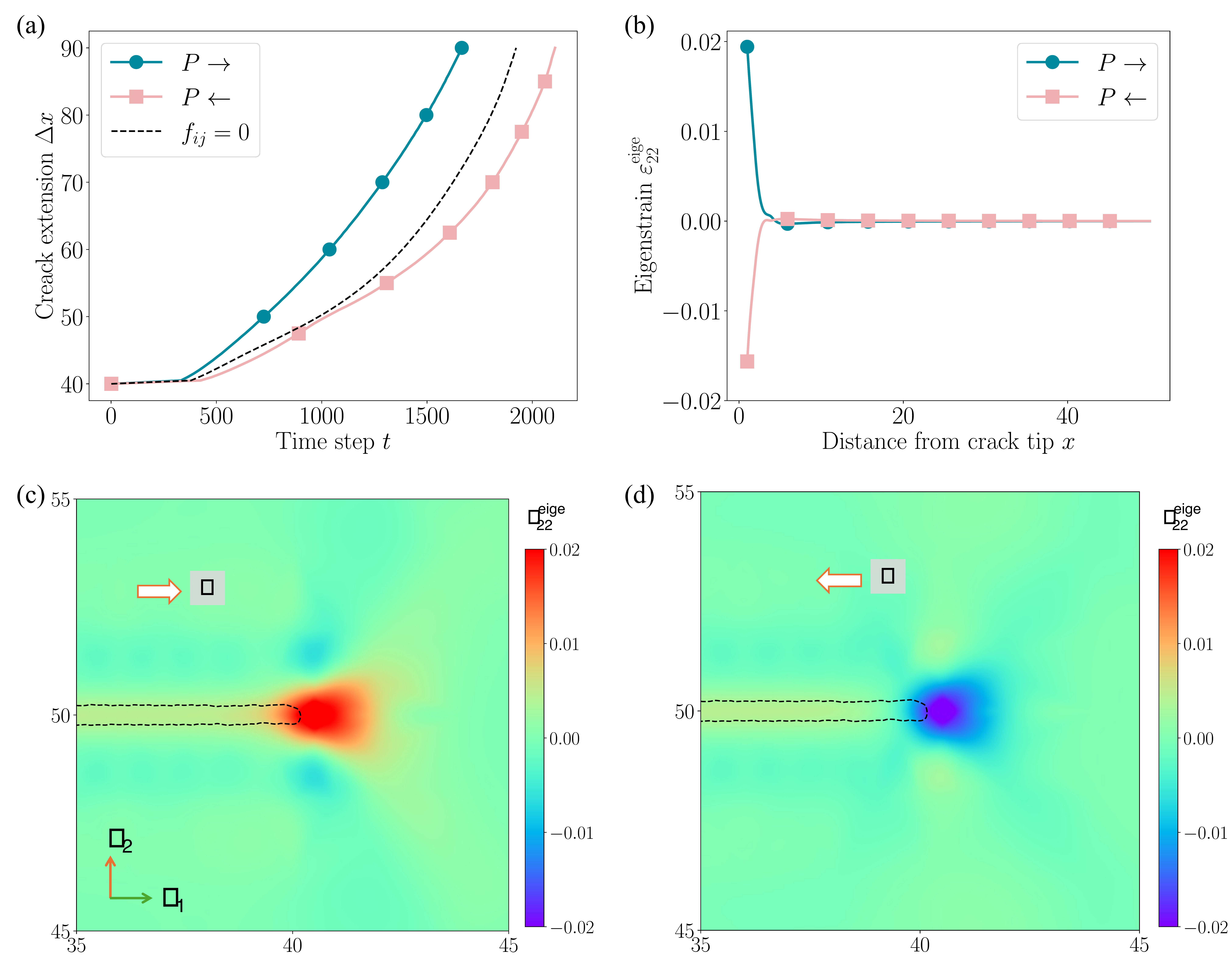}
    \caption{(a) the change in crack extension length over time step $t$. Arrows in the figure indicate the direction of initial polarization, solid lines with markers represent data considering the flexoelectric effect, and dashed lines represent data without the flexoelectric effect. (b) the change in $\varepsilon_{22}^{\mathrm{eige}}$ along $y=50$, from the crack tip to the right boundary of the plate at $t=200$. (c) and (d), the distribution of eigenstrain $\varepsilon_{22}^{\mathrm{eige}}$ near the crack tip, in which the arrows represent the direction of initial polarization, respectively. The dashed lines in the images mark the region of the crack ($\nu\geq0.9$). } \label{fig7}
\end{figure}

Figure \ref{fig7}(a) illustrates the crack extension $\Delta x$ over time step $t$. When considering the flexoelectric effect, if the polarization direction coincides with the direction of crack propagation, the crack extends earlier and requires a lower external load, consequently diminishing the material's fracture toughness. However, when the crack propagation direction is opposite to the polarization direction, the extension of the crack is delayed and requires a larger external load to initiate propagation, effectively increasing the material's fracture toughness. Thus, the flexoelectric effect can lead to asymmetric fracture toughness of ferroelectric materials. This numerical result is accordance with the previous experimental observation \cite{cordero-edwards_flexoelectric_2019}.

With the presence of the flexoelectric effect, the polarization gradient $\nabla \mathbf{P}$ induces an additional eigenstrain $\mathbf{\varepsilon}^{\mathrm{flex}}$ \cite{gu_phase-field_2014}:
\begin{equation}
    \varepsilon_{yy}^{\mathrm{eige}}=-F_{11}\frac{\partial P_y}{\partial y}-F_{12}\frac{\partial P_x}{\partial x},
\end{equation}
in which $F_{ij}=s_{ik}f_{kj}$, and $s_{kj}$ are the compliance coefficients. The polarization gradient induced eigenstrain $\varepsilon_{22}^{\mathrm{eige}}$ is concentrated on the crack tip, corresponding to the opposite polarization rotation on the two sides of the crack tip. As depicted in Figure \ref{fig7}(b), the polarization gradient generates a positive eigenstrain $\varepsilon_{22}^{\mathrm{eige}}$ at the crack tip due to the flexoelectric effect. This additional positive strain increases the strain concentration at the crack tip, consequently accelerating crack propagation. Conversely, when the polarization is oriented in the left direction, the eigenstrain $\varepsilon_{22}^{\mathrm{eige}}$ become negative at the crack tip, effectively producing an additional compression, as shown in Figure \ref{fig7}(d). This reduces the strain concentration at the crack tip and delays the crack propagation.

In addition, according to Hwang's research \cite{hwang_ferroelectricferroelastic_1995}, the domain switching should satisfy:
\begin{equation}
    \sigma_{ij}\Delta \varepsilon_{ij}+E_i\Delta P_i \geq 2P_S E_C,
\end{equation}
where $P_S$ and $E_C$ denote the magnitude of the spontaneous polarization and coercive field, respectively. Thus, the additional eigenstrain also impacts the formation of domain walls. In the cases where the polarization direction aligns along the right direction, a positive $\varepsilon_{22}^{\mathrm{eige}}$ facilitates the rotation of polarization and the earlier formation of domain structures. On the other hand, when the polarization direction is opposite to the crack propagation direction, a negative $\varepsilon_{22}^{\mathrm{eige}}$ requires a larger load to achieve polarization rotation and domain structure formation.

\begin{figure}[!htbp]
    \centering
    \includegraphics[width=0.7\linewidth]{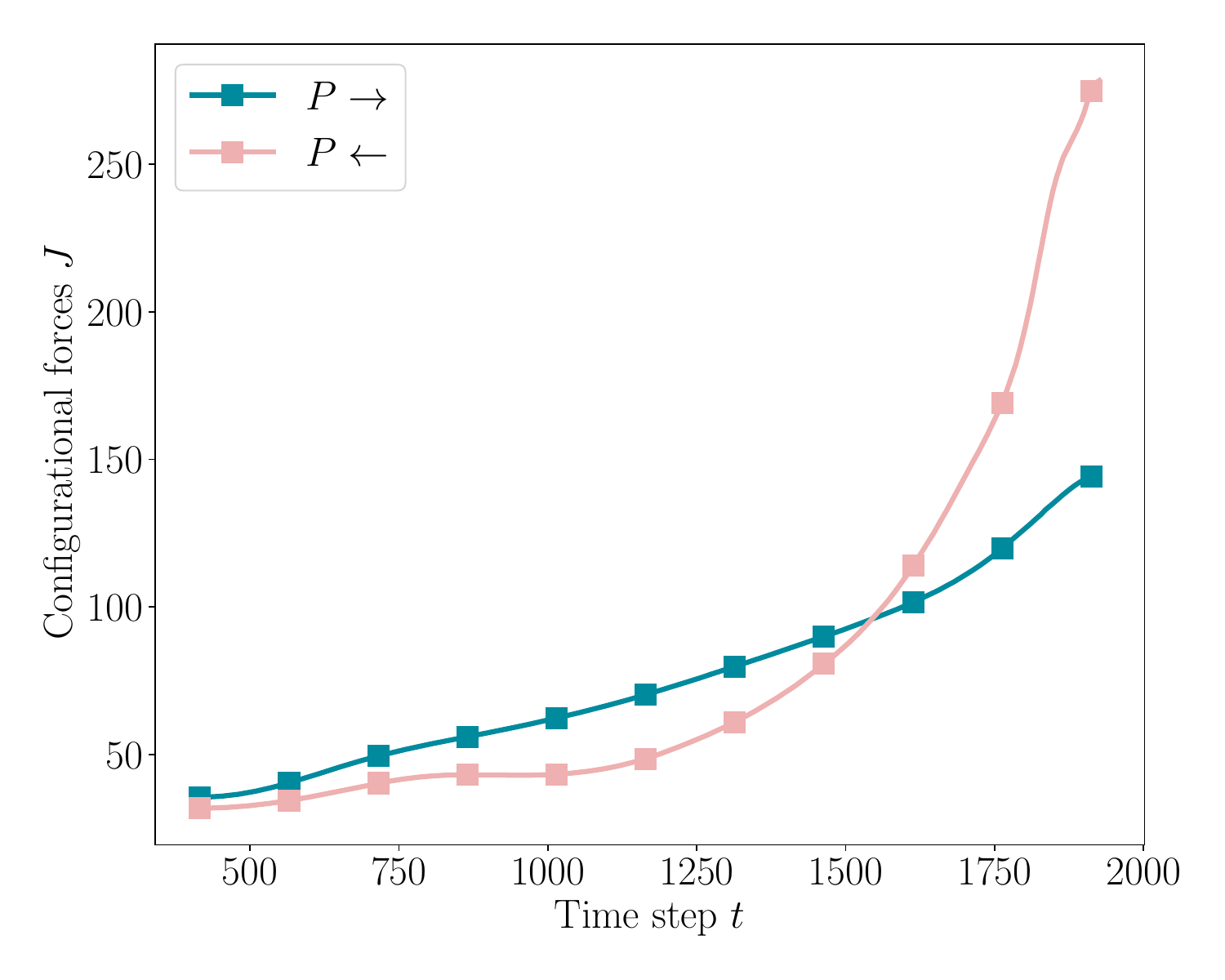}
    \caption{The variation of  fracture driving force $J$  over time step $t$.} \label{fig8}
\end{figure}

The coupling between domain evolution and flexoelectric effect can significantly influence the fracture driving force $J$. As presented in Figure \ref{fig8}, the crack driving force $J$ is higher for the initial polarization oriented in the right direction in the early stages of crack propagation. Due to the larger tensile strain at the crack tip and fewer domain walls, more energy is dedicated to driving crack growth. As the crack further extends, by the time $t^*>1500$, the cases where the polarization direction is along the left direction exhibit longer undamaged regions. Under the same displacement load $u_y$, it will endure larger stress. Additionally, as the crack extends, the domain walls begin to vanish, releasing energy and further accelerating crack propagation. Consequently, at this stage, the cases with the polarization direction aligned in the left direction have a larger driving force $J$.

\subsection{Crack perpendicular to the initial polarization direction}

\begin{figure}[!htbp]
    \centering
    \includegraphics[width=\linewidth]{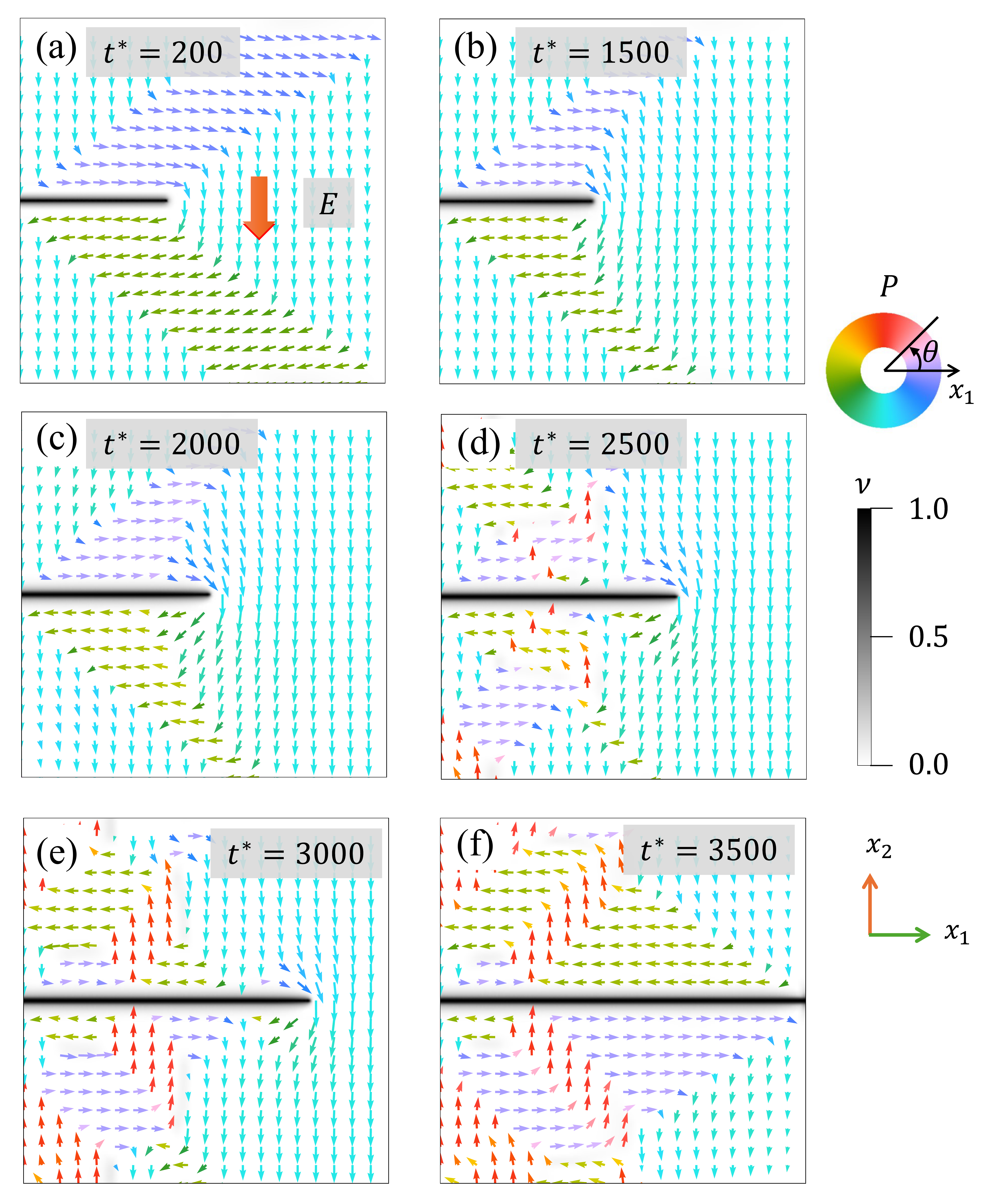}
    \caption{The dynamic progression of crack propagation and polarization switching at time step $t^*=200$, $t^*=1500$, $t^*=2000$, $t^*=2500$, $t^*=3000$ and $t^*=3500$. The grayscale maps illustrate the spatial distribution of the fracture order parameter $\nu$. The arrows indicate the orientation and magnitude of the polarization vectors. The flexoelectric coefficients $f_{ij}=0$ is applied in this figure. } \label{fig9}
\end{figure}

Figure \ref{fig9} illustrates the temporal evolution of the polarization domain structure and crack propagation in ferroelectric material. Here, the flexoelectric effect is neglected, and the electric field $E_2^*=-1$ is applied to the ferroelectric material. As shown in Figure \ref{fig9}(a), on either side of the crack, the polarization rotates $90^\circ$ in counterclockwise and clockwise direction, respectively. This results in the formation of two switched zones with forward wings structure in the material. Along the side of these wings, $90^\circ$ domain walls are formed. The tips of the wings extend to the right boundary of the plate, effectively segmenting the unswitched area into three triangular zones. Distinct from cases where the crack is aligned parallel to the initial polarization direction, domain structures are already present within the plate before crack propagation. According to Eq.\eqref{eq5}, the polarization component $P_2$ can induce a positive spontaneous strain $\varepsilon^0_{22}$.  This fundamental relationship suggests that the initial displacement load $u_2=0$ effectively introduces an additional compressive force. Besides, due to the electrical impermeability of the crack, the electric field continuity across the pre-existing  crack is disrupted. Hence, domain walls are established even prior to crack propagation.

As $u_2$ increases, the crack begins to extend. Under the influences of $u_2$ and $E_2$, the polarization in the switched zones gradually rotates to align with the direction of the electric field. Consequently, the forward wing structure progressively diminishes, as shown in Figure \ref{fig9}(b)-(c). At $t^*=2500$, the two forward wings structure has completely vanished. At this point, the polarization distribution within the material is divided into two parts by the crack tip. On the right side of the plate, where the material is intact, the polarization remains almost a single-domain state, with polarization aligned along the electric field. On the left side, an antisymmetric domain structure forms on both sides of the crack, which comprises $180\circ$ and $90^\circ$ domain walls, as presented in Figure \ref{fig9}. As $u_y$ continues to increase, the crack further extends. Two continuous polarization vortices form on either side of the crack plane, which is listed in Figure \ref{fig9}(e). After fully fractured, the plate does not revert to a single-domain state. Instead, a flux closure structure is formed on both sides of the crack, as shown in Figure \ref{fig9}{f}. Given that the crack is electrically impermeable, the areas of electrical discontinuity increase as the crack extends. Once the crack has fully extended, the regions above and below the crack effectively become open-circuited. Consequently, under the influence of external forces, the domain structures no longer revert to a single-domain state. The complete spatial and temporal evolutions of the domain structure and crack path are presented in Video S5.

\begin{figure}[!htbp]
    \centering
    \includegraphics[width=\linewidth]{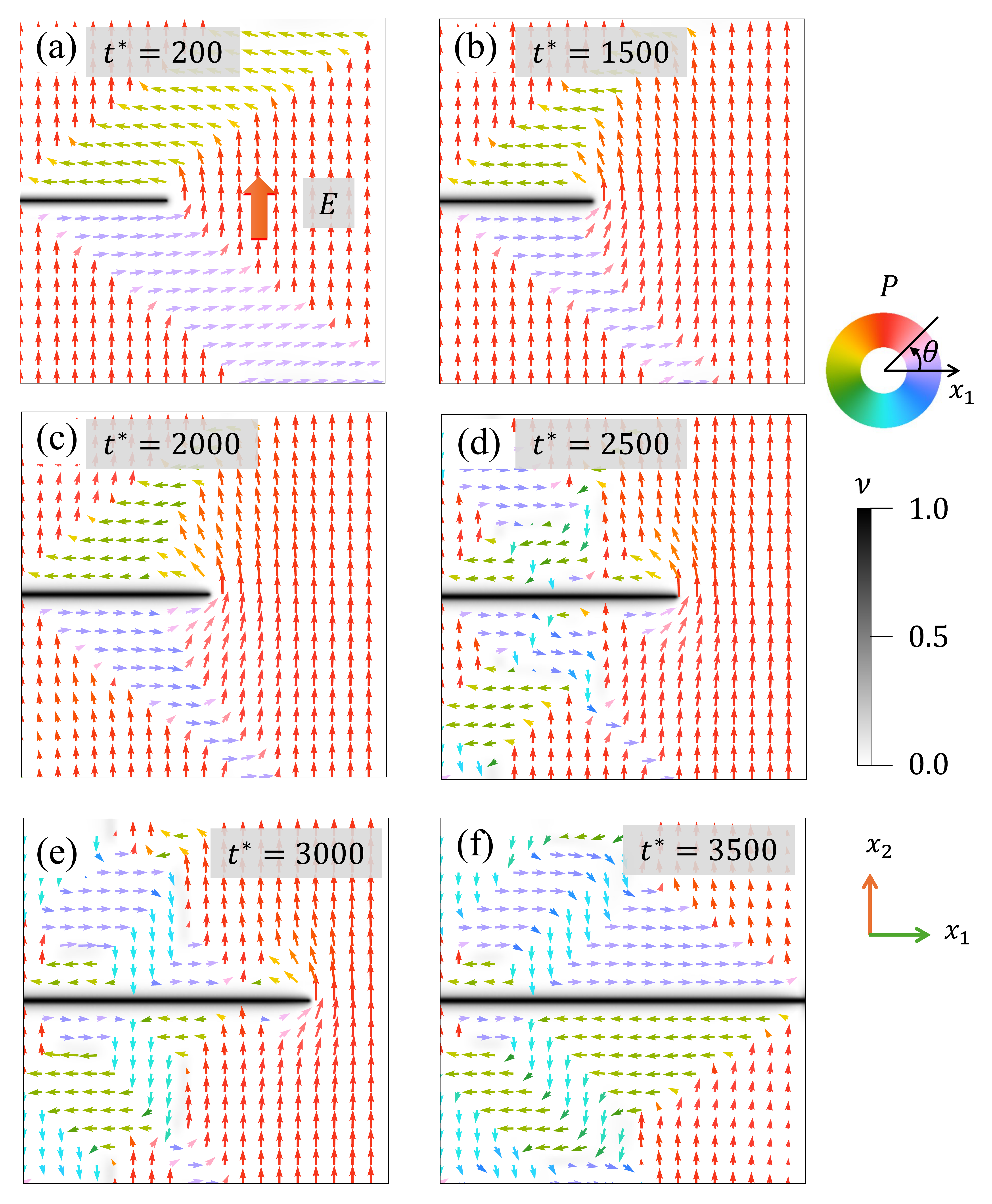}
    \caption{Morphologies of the crack and polarization at moments $t^=200$, $t^*=1500$, $t^*=2000$, $t^*=2500$, $t^*=3000$, and $t^*=3500$ under the applied electric field $E^*_2=1$, respectively. The flexoelectric effect is neglected here.} \label{fig10}
\end{figure}

Without the flexoelectric effect, the polarization evolution and crack propagation under an applied electric field of $E_2^*=1$ are identical to  those under $E_2^*=-1$. The only difference is that the polarization directions in the corresponding regions are opposite at each time step, as illustrated in Figure \ref{fig10}(a)-(f). Video S6 shows the complete evolution of the domain structure and crack path.

\begin{figure}[!htbp]
    \centering
    \includegraphics[width=0.7\linewidth]{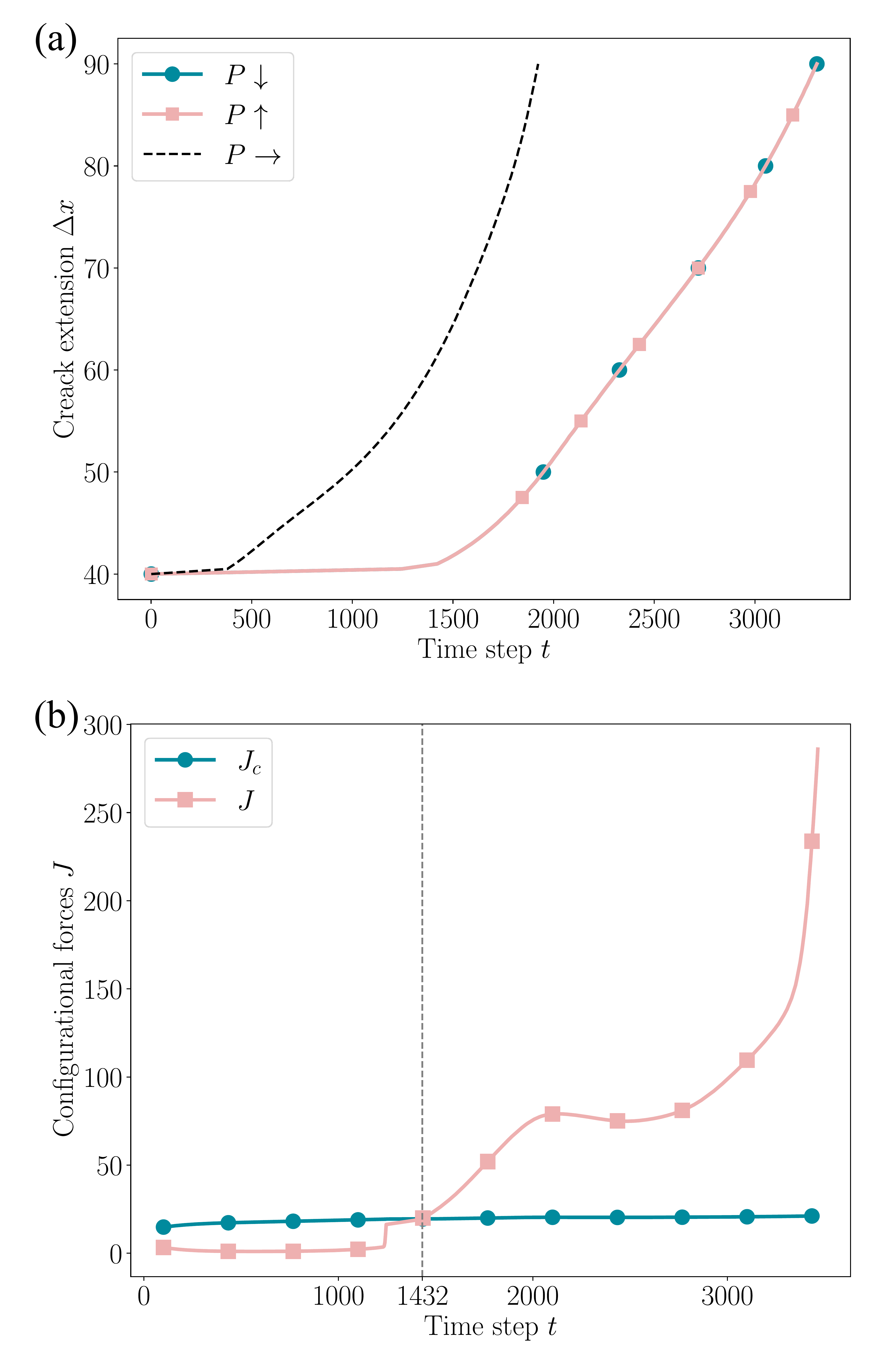}
    \caption{(a) The variation of crack extension $\Delta x$ with respect to time step $t$. The arrows in the legend denote the initial direction of polarization. (b) The change of the crack driving force $J$ and local fracture toughness $J_c$ over time step $t$. Data are collected from the cases in Figure \ref{fig9}.} \label{fig11}
\end{figure}

Figure \ref{fig11}(a) illustrates the evolution of crack extension $\Delta x$ over time $t$. Regardless of whether the initial polarization is along the upward or downward direction, the crack extends at the same rate. On the other hand, the crack extension is considerably postponed when the initial polarization direction is perpendicular to the crack. This delay arises due to the positive spontaneous strain $\varepsilon_{22}^0$ induced by the polarization component $P_2$, which places the plate in a compressed state under the $u_2=0$ boundary condition. Conversely, when the polarization is aligned along the $x_1$-axis, the resultant negative $\varepsilon_{22}^0$ will induce tension under the same boundary condition, thereby facilitating earlier crack extension.

The changes in the driving force $J$ and fracture toughness $J_c$ over time step $t$ are displayed in Figure \ref{fig11}(b). Similar to the results in Figure \ref{fig4}(b), $J_c$ remains stable around 20, which is close to the critical fracture toughness $G_c$. The crack begins to extend when $J$ exceeds $J_c$ at $t^*=1432$. During the period $t^*=1432$ and $t^*=2000$, $J$ exhibits a monotonic increase. During this period, the polarization-switched zone shrinks, releasing domain wall energy and accelerating the crack extension. After $t^*=2000$, corresponding to the formation of new domain walls, which consume part of the input energy. Hence, $J$ slightly decreases and slow down the crack extension. As $u_2$ continues to increase, the domain structure gradually stabilizes, and the energy input continues to rise. Consequently, $J$ begins to increase rapidly until fully fracture.

\begin{figure}[!htbp]
    \centering
    \includegraphics[width=\linewidth]{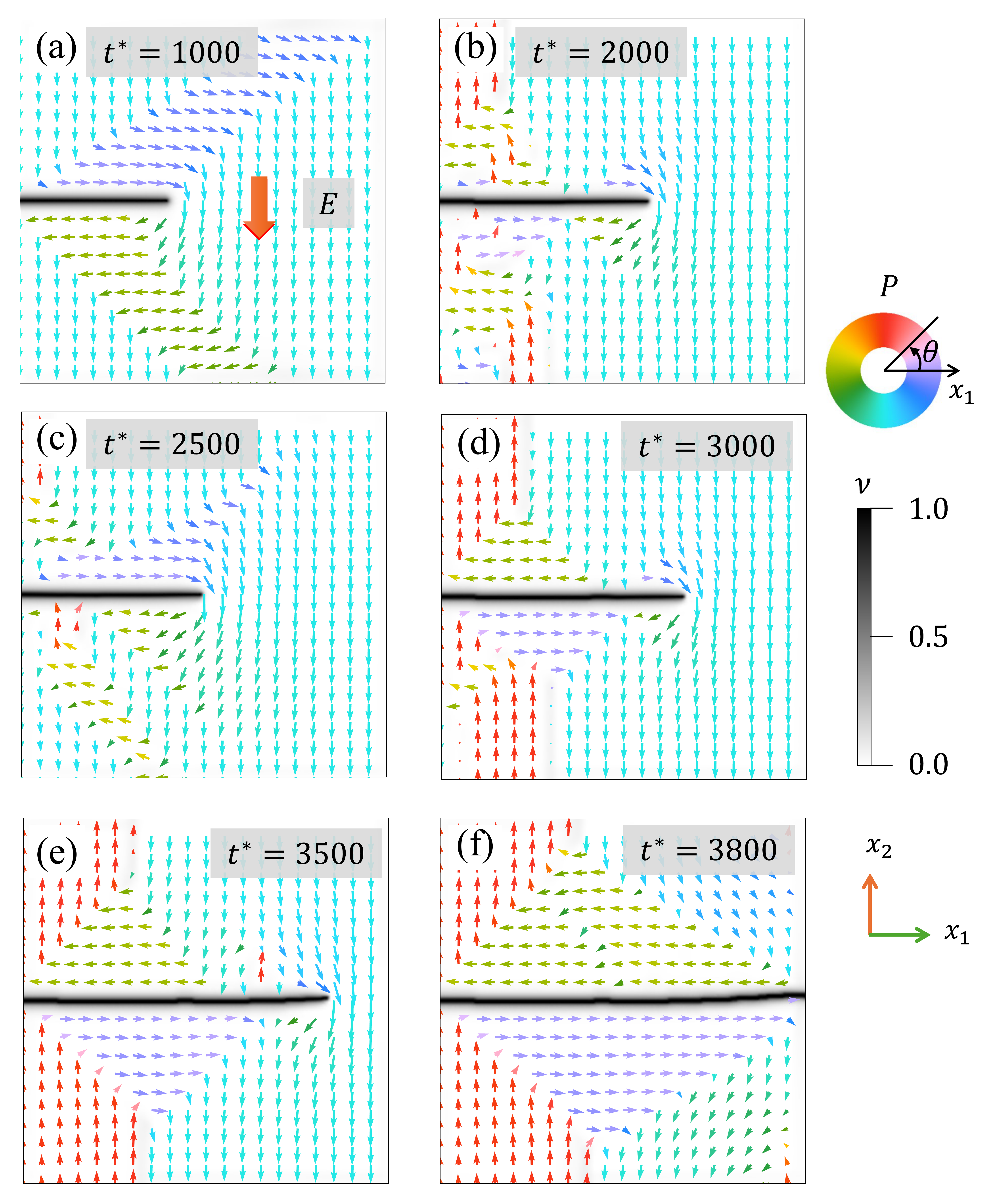}
    \caption{With the consideration of flexoelectric effect, the evolution of polarization domain and the crack extension at moments $t^*=1000$, $t^*=2000$, $t^*=2500$, $t^*=3000$, $t^*=3500$, and $t^*=3800$ under the applied electric field $E^*_y=-1$, respectively.} \label{fig12}
\end{figure}

When the initial polarization direction is perpendicular to the crack, the flexoelectric effect also influences the crack path and the evolution of domain structures during crack propagation. Figure \ref{fig12} illustrates the temporal evolution of domain structure and crack path under an applied electric field $E_2^*=-1$, taking into account the flexoelectric effect. As shown in Figure \ref{fig12}(a), before the crack begins to extend, $90^\circ$ rotation of polarization still occur on either side of the crack, which forms two forward wings structure. However, the area of the switched zone in the upper part is significantly larger than the lower counterpart, resulting in an asymmetrical domain structure. As $u_y$ increases, the crack begins to extend and the forward wing structure disappears. With the ongoing extension of the crack, the domain structures on both sides gradually evolve into a flux-closure domain structure. Additionally, during this evolution, the domain structures on either side of the crack do not maintain symmetry, as depicted in Figures \ref{fig12}(b)-(e). After complete fracture, each side of the crack forms a complete flux-closure domain, with the upper half oriented clockwise and the lower half counterclockwise. Moreover, the flexoelectric effect induces a slight upward deflection in the crack path, which no longer remains straight. The complete spatial and temporal evolution of the domain structure and crack path is presented in Video S7.
 
\begin{figure}[!htbp]
    \centering
    \includegraphics[width=\linewidth]{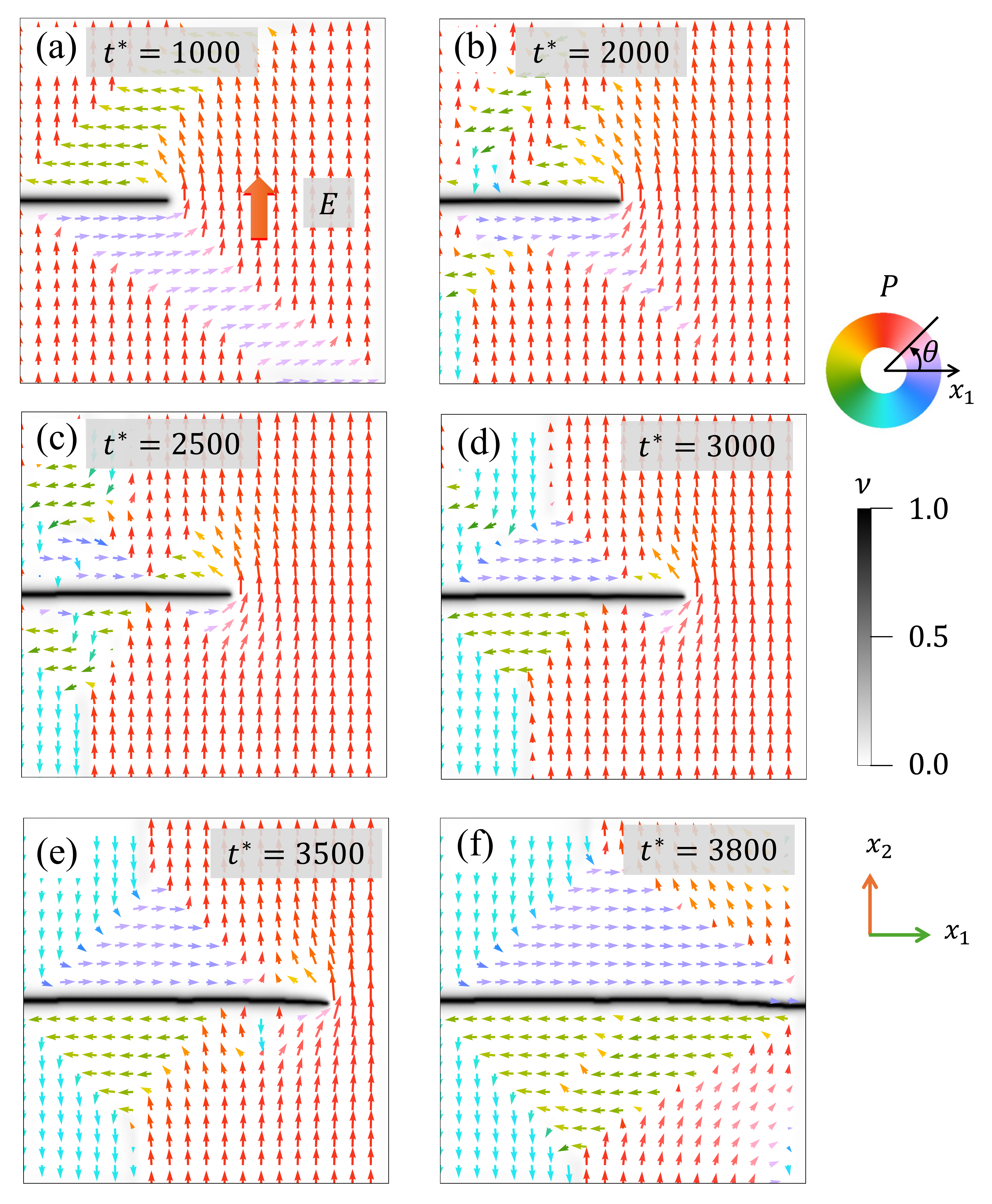}
    \caption{Evolution of polarization domain and the crack extension at moments $t^*=1000$, $t^*=2000$, $t^*=2500$, $t^*=3000$, $t^*=3500$, and $t^*=3800$ under the applied electric field $E^*_y=1$, respectively. The flexoelectric effect is considered in this case.} \label{fig13}
\end{figure}

On the other hand, when polarization is oriented in the upward direction (applied electric field $E^*_2=1$), the temporal evolution of the crack path and domain structures over time is depicted in Figure \ref{fig13}. Compared to the results listed in Figure \ref{fig12}, the polarization direction at corresponding locations is reversed by $180^\circ$.  Additionally, the crack propagation path exhibits a downward deflection. The evolution of the domain structures and the crack extension remain consistent with those shown in Figure \ref{fig12}. Video S8 displays the complete spatial and temporal evolution of the domain structure and crack path.

\begin{figure}[!htbp]
    \centering
    \includegraphics[width=\linewidth]{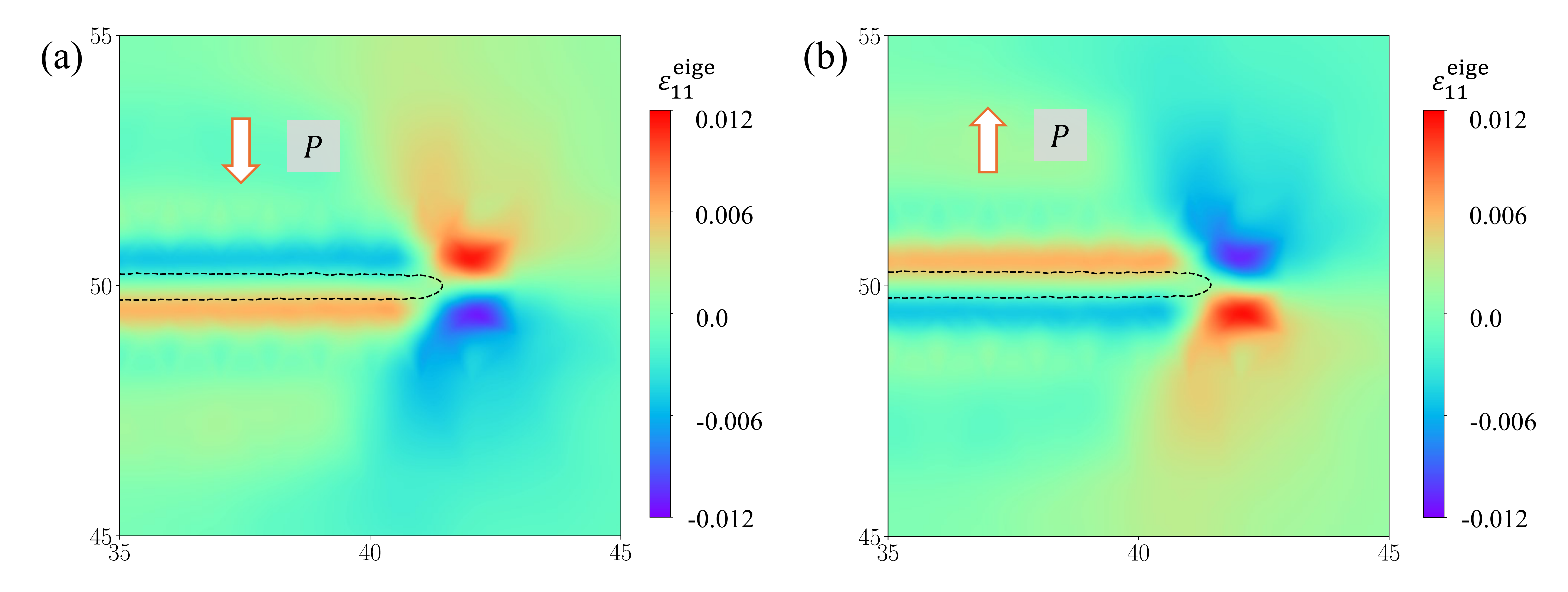}
    \caption{Local distribution of eigenstrain $\varepsilon_{11}^{\mathrm{eige}}$ near the crack tip, for (a) initial polarization along the negative $x_2$-direction and (b) initial polarization along the positive $x_2$-direction. The dashed lines in the images mark the region of the crack ($\nu\geq0.9$)} \label{fig14}
\end{figure}

With the contribution of the flexoelectric effect, the deflection of crack path can also be attributed to the eigenstrain induced by polarization gradients. Figures \ref{fig14} show the distribution of the eigenstrain $\varepsilon_{11}^{\mathrm{eige}}$ near the crack tip at $t^*=1500$. When the initial polarization is downward, additional tensile strain develops in the upper right region of the crack tip, while the lower right region experiences additional compressive strain. Consequently, $\varepsilon_{11}^{\mathrm{eige}}$ accelerates the failure of the material in the upper right region of the crack tip, while providing shielding to the lower right region. In contrast, for the   initial polarization aligned upward, the upper right area of the crack tip experiences additional compressive strain, while additional tensile strain occurs in the lower right region. Therefore, the lower right region of the crack tip is more likely to fracture, leading to the downward deflection of the crack path.

\begin{figure}[!htbp]
    \centering
    \includegraphics[width=0.7\linewidth]{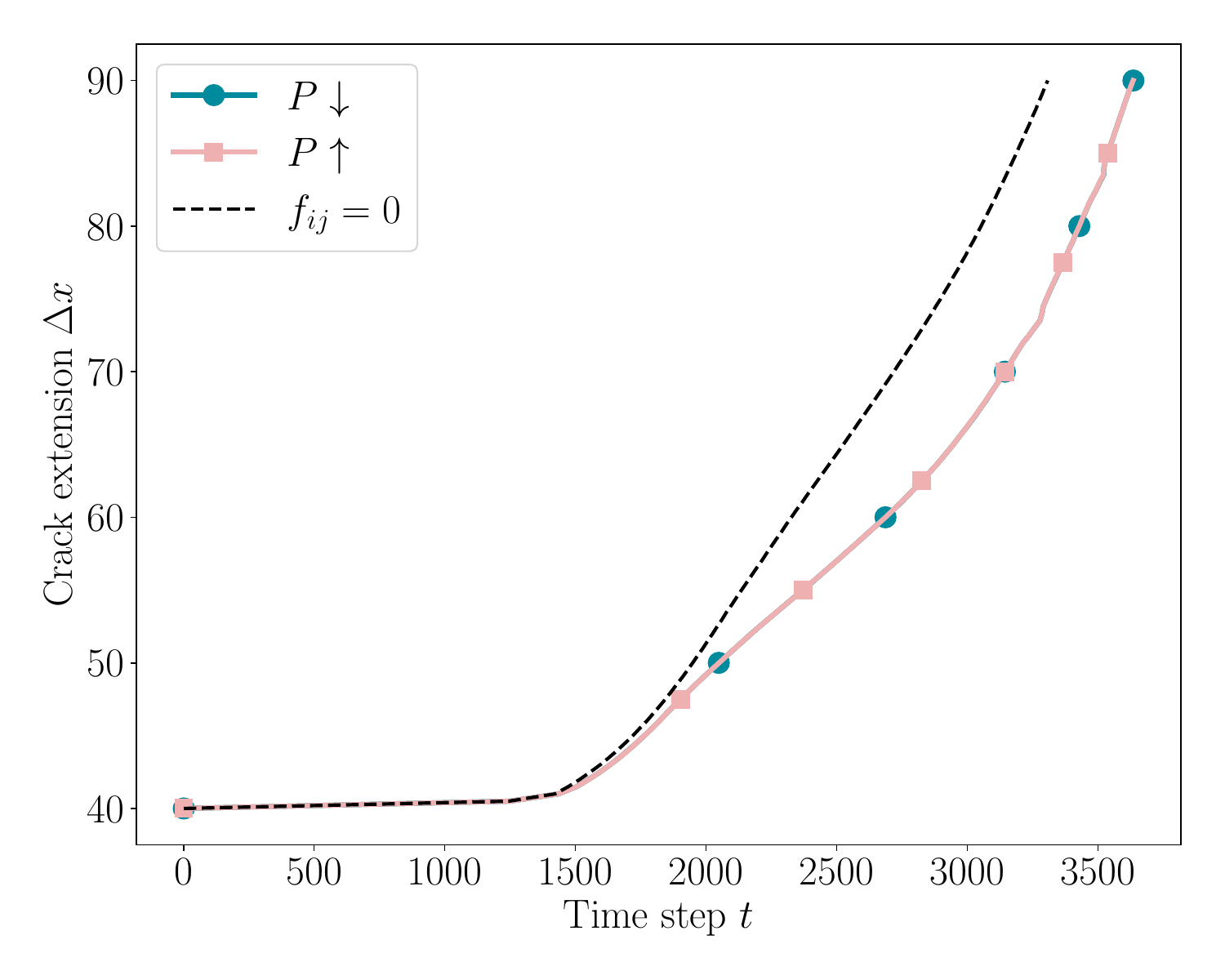}
    \caption{Crack extension $\Delta x$ over time step $t$. The solid lines represent the cases considering the flexoelectric effect, while the dashed line represents the cases without considering the flexoelectric effect. The upward and downward arrows in the legend indicate the initial polarization direction.} \label{fig15}
\end{figure}

Figure \ref{fig15} presents the crack extension $\Delta x$ over time step $t$. With the introduction of the flexoelectric effect, the onset of crack propagation is delayed. In contrast to cases where the crack is parallel to the initial polarization direction, the crack propagation rate remains identical regardless of whether the initial polarization is directed upward or downward. These simulation results are  consistent with experimental observations, which indicate that the flexoelectric effect does not alter fracture toughness when the polarization is perpendicular to the crack \cite{cordero-edwards_flexoelectric_2019}.

\begin{figure}[!htbp]
    \centering
    \includegraphics[width=0.7\linewidth]{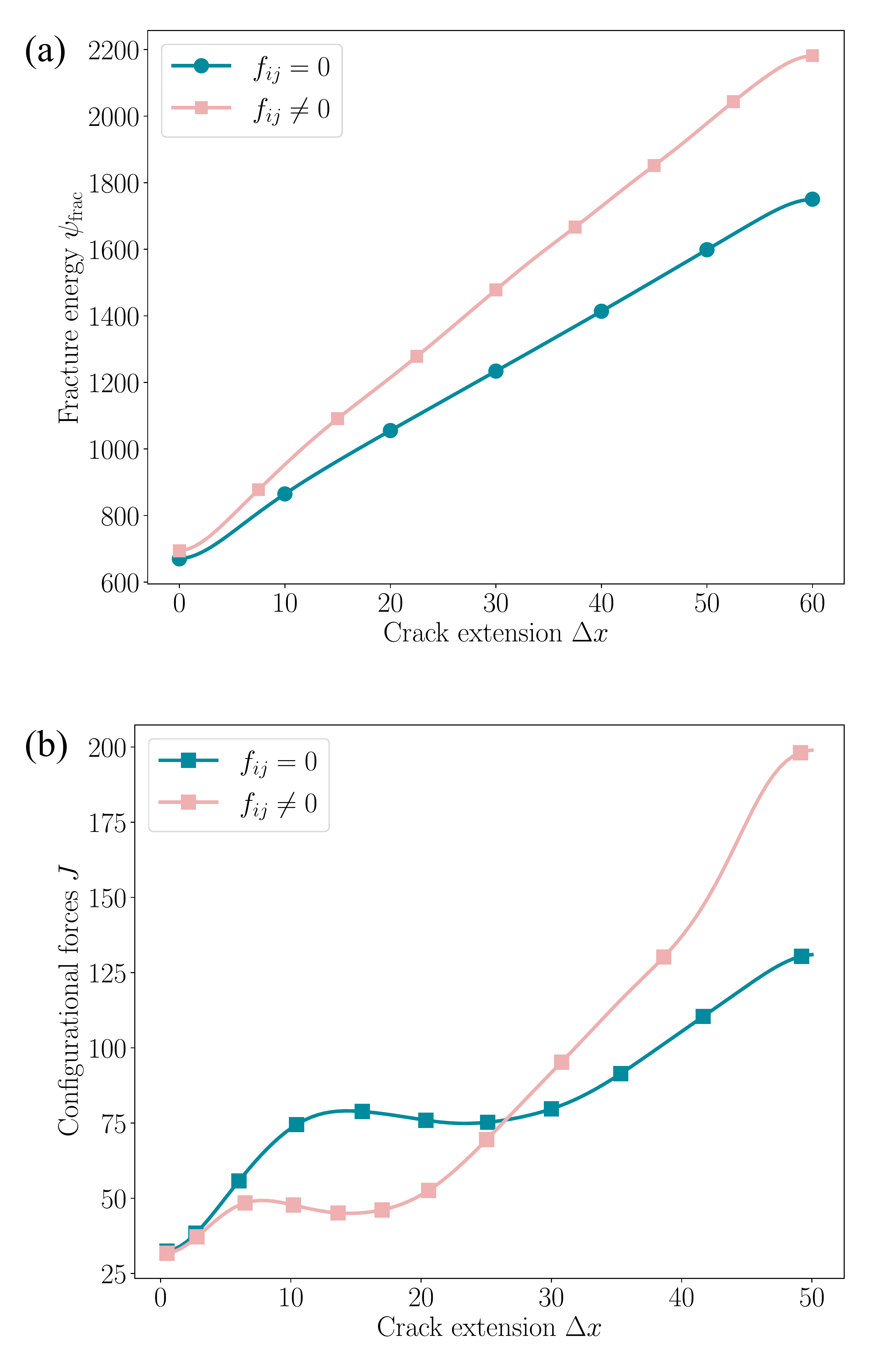}
    \caption{(a) fracture energy $\psi_\mathrm{frac}$ and (b) configurational force $J$ as functions of crack extension $\Delta x$. The data are from Figures \ref{fig9} and \ref{fig12} respectively.} \label{fig16}
\end{figure}

According to Eq.\ref{eq9}, the flexoelectric effect causes a deflection in the crack propagation path, which increases the overall crack length. Figure \ref{fig16} compares the fracture energy $\psi_\mathrm{frac}$ as a function of crack extension $\Delta x$ for cases with  or without the flexoelectric effects. Due to the presence of pre-existing cracks, the $\psi_\mathrm{frac}$ values for both cases are similar at  the initial state. As the crack extends, both cases show a monotonic increase in $\psi_\mathrm{frac}$. However, with the consideration of the flexoelectric effect, the deflection of the crack path leads to a more rapid increase in $\psi_\mathrm{frac}$, indicating that more energy is expended on crack growth.

The difference in the fracture energy can also influence the crack driving force $J$. As presented in Figure \ref{fig16}(b), during the early stages of crack extension, the crack driving force becomes larger when the flexoelectric effect is neglected. Hence, the crack propagation rate is faster without the flexoelectric effect. This is consistent with the results that more energy is required to form a deflected crack path with the flexoelectric effect. Throughout the crack propagation, the intact region of the material is longer at the same time step compared with the case that considering the flexoelectric effect. As a result, it also stands a larger load. Consequently, as the crack extension, the difference in input energy gradually compensates for the difference in fracture energy. Therefore, in the latter stages of crack propagation with flexoelectric effect, the crack driving force is higher, and the crack extends more rapidly. 

\section{Concluding remarks}\label{Section5}
In this study, a phase-field fracture model for ferroelectric materials that incorporates the flexoelectric effect is established. The model uses isogeometric analysis method with PHT-splines to realize the local refinement near the crack. The crack propagation process in ferroelectric materials with electrically impermeable crack conditions is simulated. The results indicate that the crack propagation in ferroelectric materials depends on the polarization direction with the consideration of flexoelectric effect. When the crack propagation direction aligns with the polarization direction, the crack extends the fastest. In contrast, the crack extension is decelerated when the propagation direction opposes the polarization direction. When the initial polarization is perpendicular to the crack, the flexoelectric effect can induce a deflection in the crack path, which is opposite to the initial polarization direction. The underlying mechanism can be attributed to the eigenstrain induced by the flexoelectric effect.

The numerical results are consistent with previous experimental observations  and  provide a phenomenological explanation for the anisotropic effects during fracture in ferroelectric materials. From this work, some further researches can also be studied. For example, this study used an electrically impermeable crack. The impact of different electrical boundary conditions on crack propagation still needs investigation. Additionally, a two-dimensional model was used to simplify calculations. However, for materials like PMN-PT, a two-dimensional model cannot accurately describe their domain structures. Therefore, a three-dimensional model is still needed to study the fracture process.

\section*{Acknowledgments}
This work is financially supported by the National Nature Science Foundation of China (Grant No. 12202370, 12102387, and 12192214), Natural Science Foundation of Sichuan Province of China (No. 24NSFSC4777) and New Interdisciplinary Cultivation Fund of Southwest Jiaotong University (Grant No. 2682022KJ050).

\appendix
\section{Residual and the tangent matrices}
The weak form of governing equation can be expressed in matrix form:
\begin{equation}\label{eq23}
    \begin{split}
        &\int_\Omega\bigg[ \delta\mathbf{\varepsilon}^T\mathbf{c}\mathbf{\varepsilon}-\delta\mathbf{\varepsilon}^T\mathbf{Q}\mathbf{P}+\delta\mathbf{\varepsilon}^T\mathbf{f}\mathbf{\xi}-\delta\mathbf{\gamma}^T\mathbf{f'}\mathbf{P}-\delta\mathbf{E}^T\mathbf{k}\mathbf{E}+\delta\mathbf{P}^T\mathbf{\alpha}\mathbf{P}\\
        &-\delta\mathbf{P}^T\mathbf{Q'}\mathbf{\varepsilon}-\delta\mathbf{P}^T{\mathbf{f}'}^T\mathbf{\gamma}+\frac{1}{L_P}\delta\mathbf{P}^T \mathbf{\dot{P}} +\delta\mathbf{\xi}^T\mathbf{G}\mathbf{\xi}+\delta\mathbf{\xi}^T\mathbf{f}\mathbf{\varepsilon}\\
        &+\delta v\cdot\left( \frac{1}{L_\nu} \dot{v}+ \delta v\frac{\partial F'}{\partial v}+G_c\frac{v-1}{2k} \right)+4\delta v_{,i}G_ckv_{,i} \bigg] \mathrm{d} V\\
        =&\int_{\partial \Omega} \left( \delta \mathbf{u}^T\mathbf{\tau}-\delta \mathbf{\varepsilon}^T\mathbf{M}-\delta \phi \cdot \omega + \delta\mathbf{P}^T\mathbf{\pi}+\delta v\zeta\right)\mathrm{d}S
    \end{split}
\end{equation}
with the material coefficient matrices
\begin{equation}
    \begin{split}
        &
        {\mathbf{c}}=g(\nu)\begin{bmatrix}
            c_{11} & c_{12} & 0 \\
            c_{12} & c_{11} & 0 \\
            0 & 0 & c_{44}
        \end{bmatrix},\;
        {\mathbf{Q}(P_i)}=g(\nu)\begin{bmatrix}
            q_{11}P_1 & q_{12}P_2 \\
            q_{12}P_1 & q_{11}P_2 \\
            0 &  q_{44}P_1
        \end{bmatrix},\;\\
        &
        {\mathbf{f}}=\frac{1}{2}g(\nu)\begin{bmatrix}
            f_{11} & f_{12} & 0 & 0 \\
            f_{12} & f_{11} & 0 & 0 \\
            0 & 0 & f_{44} & f_{44}
        \end{bmatrix}, \;
        {\mathbf{f}'}=\frac{1}{2}g(\nu)\begin{bmatrix}
            f_{11} & 0 \\
            0 & f_{12} \\
            f_{12} & 0 \\
            0 & f_{11} \\
            0 & f_{44} \\
            f_{44} & 0
        \end{bmatrix},\\
        &
        {\mathbf{\kappa}}=g(\nu)\begin{bmatrix}
            \kappa & 0 \\
            0 & \kappa
        \end{bmatrix}\; \text{(electrical impermeable)},\;
        {\mathbf{\kappa}}=\begin{bmatrix}
            \kappa & 0 \\
            0 & \kappa
        \end{bmatrix}\; \text{(electrical permeable)},\; \\
        &
        \mathbf{\alpha}(P_i)=\begin{bmatrix}
             2\alpha_1+4\alpha_{11}P_1^2+2\alpha_{12}P_2^2+6\alpha_{111}P_1^4 & {}\\
             +\alpha_{112}(2P_2^4+4P_1^2P_2^2)+8\alpha_{1111}P_1^6  & 0\\
             +\alpha_{1112}(6P_1^4P_2^2+2P_2^6)+4\alpha_{1122}P_1^2P_2^4 & {} \\
             {}& 2\alpha_1+4\alpha_{11}P_2^2+2\alpha_{12}P_1^2+6\alpha_{111}P_2^4 \\
             0 & +\alpha_{112}(2P_1^4+4P_1^2P_2^2)+8\alpha_{1111}P_2^6\\
             {}& +\alpha_{1112}(6P_2^4P_1^2+2P_1^6)+4\alpha_{1122}P_2^2P_1^4 
        \end{bmatrix}
        \\
        & 
        {\mathbf{Q}'(P_i)}=g(\nu)\begin{bmatrix}
            2q_{11}P_1 & 2q_{12}P_1 & q_{44}P_2 \\
            2q_{12}P_2 & 2q_{11}P_2 & q_{44}P_1
        \end{bmatrix}, \\
        &
        {\mathbf{G}}=g(\nu)\begin{bmatrix}
            G_{11} & G_{12} & 0 & 0 \\
            G_{12} & G_{11} & 0 & 0 \\
            0 & 0 & G_{44}+G_{44}' & G_{44}-G_{44}' \\
            0 & 0 & G_{44}-G_{44}' & G_{44}+G_{44}' \\
        \end{bmatrix}, \;
        \mathbf{G}_\nu=\begin{bmatrix}
            4G_ck & 0 \\
            0 & 4G_ck
        \end{bmatrix}.
    \end{split}
\end{equation}

The damage model presented in the previous section is implemented by the 2D isogeometric analysis (IGA). In the implementation the displacement $\mathbf{u}$, the electric potential $\phi$, and the spontaneous polarization $\mathbf{P}$, and the damage variable $v$ are taken as the nodal degrees of freedom. Thus, each node has six degrees of freedom, namely
\begin{equation}
    \mathbf{d}^I=\left[ u_1^I\quad u_2^I\quad \phi\quad P_1^I\quad P_2^I\quad v \right]^T,
\end{equation}
where the superscript $I$ indicates the variable at element node. 
Substituting Eqs.\eqref{eq40}-\eqref{HB} into Eq.\eqref{eq23}, the weak form of governing equation can be rewritten as 
\begin{equation}
\begin{split}
        &\begin{bmatrix}
            \mathbf{K}_{uu} & 0 & -\mathbf{K}_{uP}+\mathbf{K}_{u\xi}-\mathbf{K}_{\gamma P} & 0 \\
            0 & -\mathbf{K}_{\phi\phi} & -\mathbf{K}_{\phi P} & 0 \\
            -\mathbf{K}_{Pu}+\mathbf{K}_{\xi u}-\mathbf{K}_{P \gamma} & -\mathbf{K}_{P\phi } & \mathbf{K}_{PP}+K'_{PP} & 0 \\
            0 & 0 & 0 & \mathbf{K}_{\nu\nu}+\mathbf{K}'_{\nu\nu}
        \end{bmatrix}
        \begin{bmatrix}
            \mathbf{u}\\ \mathbf{\phi} \\ \mathbf{P} \\ \mathbf{v}
        \end{bmatrix}^I \\
       + &
        \begin{bmatrix}
            0 & 0 & 0 & 0 \\
            0 & 0 & 0 & 0 \\
            0 & 0 & \mathbf{K}_{Pt} & 0 \\
            0 & 0 & 0 & \mathbf{K}_{\nu t} 
        \end{bmatrix}
        \begin{bmatrix}
            0\\ 0 \\ \dot{\mathbf{P}} \\ \dot{\mathbf{v}}
        \end{bmatrix}^I
        =
        \begin{bmatrix}
            \mathbf{F}_S+\mathbf{M}_S \\
            {\Omega}_S \\
            \mathbf{\Pi}_S \\
            \mathbf{\zeta}
        \end{bmatrix}
\end{split}
\end{equation}
in which 
\begin{equation}
    \begin{split}
        & \mathbf{K}_{uu}=\int_\Omega \mathbf{B}_u^T{\mathbf{c}}\mathbf{B}_u\,\mathrm{d}V;\;  
        \mathbf{K}_{uP}=\int_\Omega \mathbf{B}_u^T{\mathbf{Q}(P_i)}\mathbf{N}_P\,\mathrm{d}V; \; \\
        & \mathbf{K}_{u\xi}=\int_\Omega \mathbf{B}_u^T{\mathbf{f}}\mathbf{B}_P\,\mathrm{d}V;\;
        \mathbf{K}_{\gamma P}=\int_\Omega \mathbf{H}_u^T{\mathbf{f'}}\mathbf{N}_P\,\mathrm{d}V; \\
        &\mathbf{K}_{\phi\phi}=\int_\Omega \mathbf{B}_\phi^T{\mathbf{\kappa}}\mathbf{B}_\phi\,\mathrm{d}V; \;
        \mathbf{K}_{PP}=\int_\Omega \mathbf{N}_P^T\mathbf{\alpha}(P_i)\mathbf{N}_P\,\mathrm{d}V; \; \\
        &\mathbf{K}_{PP'}=\int_\Omega \mathbf{B}_P^T{\mathbf{G}}\mathbf{B}_P\,\mathrm{d}V; \; \mathbf{K}_{Pu}=\int_\Omega \mathbf{N}_P^T{\mathbf{Q}'(P_i)}\mathbf{B}_u\, \mathrm{d}V; \\
        &\mathbf{K}_{\nu\nu}'=\int_\Omega \mathbf{B}_\nu^T \mathbf{G}_\nu \mathbf{B}_\nu\, \text{d}V; \;
        \mathbf{K}_{\xi u}=\mathbf{K}_{u\xi}^T;\;
        \mathbf{K}_{P\gamma}=\mathbf{K}_{\gamma P}^T;\; 
         \\
        &\mathbf{F}_S=\int_{\partial \Omega} \mathbf{N}_u^T\mathbf{N}_T\,\mathrm{d}S \, \mathbf{T}^I; \;
        \mathbf{M}_S=\int_{\partial \Omega} \mathbf{B}_u^T\mathbf{N}_M\,\mathrm{d}S \, \mathbf{M}^I; \\
        & \mathbf{\Omega}_S=-\int_{\partial \Omega} \mathbf{N}_\phi^T\mathbf{N}_\omega\,\mathrm{d}S \, \mathbf{\Omega}^I;\; 
        \mathbf{\Pi}_S=\int_{\partial \Omega} \mathbf{N}_P^T\mathbf{N}_\pi \mathrm{d}S \,\mathbf{\Pi}^I; \\
        &\mathbf{\zeta}=\int_{\partial \Omega} \mathbf{N}_\nu^T \mathbf{N}_\zeta\mathrm{d}S\, \mathbf{\zeta}^I+\int_\Omega \mathbf{N}_\nu^T\frac{G_c}{l_\nu}\,\text{d}V.
    \end{split}
\end{equation}

For electrical permeable crack,
\begin{equation}
    \begin{split}
        & \mathbf{K}_{\phi P}=\int_\Omega \mathbf{B}_\phi^T\mathbf{N}_P\,\mathrm{d}V; \\
        &\mathbf{K}_{\nu\nu}=\int_\Omega \mathbf{N}_\nu^T \left[(f_\mathrm{grad}+f_\mathrm{elas}+f_\mathrm{flex})+\frac{G_c}{l_\nu}\right]\mathbf{N}_\nu \,\mathrm{d}V.
    \end{split}
\end{equation}
For electrical impermeable crack, 
\begin{equation}
    \begin{split}
        &\mathbf{K}_{\phi P}=\int_\Omega \mathbf{B}_\phi^T{\mathbf{v}}\mathbf{N}_P\,\mathrm{d}V; \\
        &\mathbf{K}_{\nu\nu}=\int_\Omega \mathbf{N}_\nu^T \left[(f_\mathrm{grad}+f_\mathrm{elas}+f_\mathrm{flex}+f_\mathrm{elec})+\frac{G_c}{l_\nu}\right]\mathbf{N}_\nu \,\mathrm{d}V .
    \end{split}
\end{equation}

Newton-Raphson method and the backward Euler method is applied here to solve the time-dependent nonlinear partial differential equations. The iteration equation in Newton-Raphson method is 
\begin{equation}
     \mathbf{d}^{n+1}=\mathbf{d}^n-\mathbf{R}^n/\mathbf{S}^n,
\end{equation}
with 
\begin{equation}
    \begin{split}
        &\mathbf{R}= \\
        &\begin{bmatrix}
            \mathbf{K}_{uu} & 0 & -\mathbf{K}_{uP}+\mathbf{K}_{u\xi}-\mathbf{K}_{\gamma P} & 0 \\
            0 & -\mathbf{K}_{\phi\phi} & \mathbf{K}_{\phi P} & 0 \\
            -\mathbf{K}_{Pu}+\mathbf{K}_{\xi u}-\mathbf{K}_{P \gamma} & -\mathbf{K}_{P\phi } & \mathbf{K}_{PP}+K'_{PP} & 0 \\
            0 & 0 & 0 & \mathbf{K}_{\nu\nu}+\mathbf{K}'_{\nu\nu}
        \end{bmatrix}
        \begin{bmatrix}
            \mathbf{u}\\ \mathbf{\phi} \\ \mathbf{P} \\ \mathbf{v}
        \end{bmatrix}^I \\
        &+ 
        \begin{bmatrix}
            0 & 0 & 0 & 0 \\
            0 & 0 & 0 & 0 \\
            0 & 0 & \mathbf{K}_{Pt} & 0 \\
            0 & 0 & 0 & \mathbf{K}_{vt} 
        \end{bmatrix}
        \begin{bmatrix}
            0\\ 0 \\ \dot{\mathbf{P}} \\ \dot{\mathbf{v}}
        \end{bmatrix}^I
        -
        \begin{bmatrix}
            \mathbf{F}_S+\mathbf{M}_S \\
            {\Omega}_S \\
            \mathbf{\Pi}_S \\
            \mathbf{\zeta}_S
        \end{bmatrix}
    \end{split}
\end{equation}
\begin{equation}
    \mathbf{S}=\begin{bmatrix}
        \mathbf{K}_{uu} & 0 & -\mathbf{K}_{uP}^S+\mathbf{K}_{u\xi}-\mathbf{K}_{\gamma P} & \mathbf{K}_{uv}^S \\
        0 & -\mathbf{K}_{\phi\phi} & -\mathbf{K}_{\phi P} & {\mathbf{k}_{\phi v}^S} \\
        -\mathbf{K}_{Pu}^S+\mathbf{K}_{\xi u}-\mathbf{K}_{P \gamma} & -\mathbf{K}_{P\phi } & \mathbf{K}_{PP}^S+\mathbf{K}'_{PP}+\mathbf{K}_{Pt}^S & \mathbf{K}_{Pv}^S \\
        \mathbf{K}_{vu}^S & {\mathbf{k}_{v\phi}^S} & \mathbf{K}_{vP}^S & \mathbf{K}_{\nu\nu}+\mathbf{K}'_{\nu\nu}+\mathbf{K}_{vt}^S
    \end{bmatrix},
\end{equation}
in which 
\begin{equation}
    \mathbf{K}_{uP}^S=\int_\Omega \mathbf{B}_u^T \mathbf{Q}'^T \mathbf{N}_P\, \text{d}V,
\end{equation}
\begin{equation}
    \mathbf{K}_{uv}^S=\int_\Omega\left(\mathbf{B}_u^T\mathbf{c}_\nu\mathbf{N}_\nu-\mathbf{B}_u^T\mathbf{Q}_\nu\mathbf{N}_\nu+\mathbf{B}_u^T\mathbf{f}_\nu\mathbf{N}_\nu-\mathbf{H}_u^T\mathbf{f}_\nu'\mathbf{N}_\nu\right)\mathrm{d}V,\\
\end{equation}
\begin{equation}
    \begin{split}
    &{\mathbf{K}_{\phi v}}=0 \text{ (electrical permeable)},\;\\
    &{\mathbf{K}_{\phi v}}=\int_\Omega \mathbf{B}_\phi \mathbf{\kappa}_\nu \mathbf{N}_\nu\, \text{d}V \text{ (electrical permeable)},\;
\end{split}
\end{equation}
\begin{equation}
    \mathbf{K}_{Pu}^S=\int_\Omega \mathbf{N}_P^T \mathbf{Q}^S\mathbf{B}_u \, \text{d}V,
\end{equation}
\begin{equation}
    \mathbf{K}_{PP}^S=\int_\Omega \mathbf{N}_P^T \mathbf{\alpha}^S \mathbf{N}_P\, \text{d}V,
\end{equation}
\begin{equation}
    \mathbf{K}_{Pt}^S=\int_\Omega \mathbf{N}_P^T \mathbf{P}_{i,t}^S \mathbf{N}_P\, \text{d}V,
\end{equation}
\begin{equation}
    \mathbf{K}_{Pv}^S=\int_\Omega (-\mathbf{N}_P \mathbf{Q}'\mathbf{N}_\nu+\mathbf{B}_P\mathbf{f}_\nu^T\mathbf{N}_\nu-\mathbf{N}_P\mathbf{f}'^T\mathbf{N}_\nu)\, \text{d}V
\end{equation}
\begin{equation}
    \mathbf{K}_{\nu t}^S=\int_\Omega \mathbf{N}_\nu^T \mathbf{v}_{,t}^S \mathbf{N}_\nu\, \text{d}V.
\end{equation}

And the corresponding stiffness matrixes
\begin{equation}
\begin{split}
        &
        \mathbf{c}_\nu=g'(\nu)\begin{bmatrix}
            c_{11}\varepsilon_{11}+c_{12}\varepsilon_{22} \\
            c_{12}\varepsilon_{11}+c_{11}\varepsilon_{22} \\
            2c_{44}\varepsilon_{12}
        \end{bmatrix},\;
        \mathbf{Q}_\nu=g'(\nu)\begin{bmatrix}
            q_{11}P_1^2+q_{12}P_2^2 \\
            q_{12}P_1^2+q_{11}P_2^2 \\
            q_{44}P_1P_2 \\
        \end{bmatrix},\; \\
        &
        \mathbf{f}_\nu=\frac{1}{2}g'(\nu)\begin{bmatrix}
            f_{11}P_{1,1}+f_{12}P_{2,2} \\
            f_{12}P_{1,1}+f_{11}P_{2,2} \\
            f_{44}(P_{1,2}+P_{2,1}) \\
        \end{bmatrix},\;        
        \mathbf{f}_\nu'=\frac{1}{2}g'(\nu)\begin{bmatrix}
            f_{11}P_1 \\
            f_{12}P_2 \\
            f_{12}P_1 \\
            f_{11}P_2 \\
            f_{44}P_2 \\
            f_{44}P_1
        \end{bmatrix}, \\
        &
        \mathbf{\kappa}_\nu=g'(\nu)\begin{bmatrix}
            -\kappa E_1 - P_1 \\
            -\kappa E_2 - P_2 
        \end{bmatrix},
        \mathbf{Q}^S=g(\nu)\begin{bmatrix}
            2Q_{11}P_1 & 2Q_{12}P_1 & Q_{44}P_2 \\
            2Q_{12}P_2 & 2Q_{11}P_2 & Q_{44}P_1
        \end{bmatrix}, \\
        &
        \mathbf{P}_{i,t}^S=\begin{bmatrix}
            \dfrac{1}{\Delta t} & 0 \\
            0 & \dfrac{1}{\Delta t} \\
        \end{bmatrix}, \;
        \mathbf{v}_{,t}^S=\begin{bmatrix}
            \dfrac{1}{\Delta t} & 0 \\
            0 & \dfrac{1}{\Delta t} \\
        \end{bmatrix}, \\
        &
        \mathbf{\alpha}^S=\begin{bmatrix}
            2\alpha_1 +12 \alpha_{11}P_1^2+2\alpha_{12}P_2^2+30\alpha_{111}P_1^4 & {4\alpha_{12}P_1P_2+8\alpha_{112}(P_1P_2^3+P_1^3P_2)} \\
            +\alpha_{112}(2P_2^4+12P_1^2P_2^2)+56\alpha_{1111}P_1^6 & +12\alpha_{1112}(P_1^5P_2+P_1P_2^5) \\
            +\alpha_{1112}(30P_1^4P_2^2+2P_2^6)+12\alpha_{1122}P_1^2P_2^4 & {+16\alpha_{1122}P_1^3P_2^3} \\
            {}&{}\\
            {4\alpha_{12}P_1P_2+8\alpha_{112}(P_1P_2^3+P_1^3P_2)} & 2\alpha_1 +12 \alpha_{11}P_2^2+2\alpha_{12}P_1^2+30\alpha_{111}P_2^4 \\
            +12\alpha_{1112}(P_1^5P_2+P_1P_2^5) & +\alpha_{112}(2P_1^4+12P_1^2P_2^2)+56\alpha_{1111}P_2^6 \\
            {+16\alpha_{1122}P_1^3P_2^3} & +\alpha_{1112}(30P_2^4P_1^2+2P_1^6)+12\alpha_{1122}P_2^2P_1^4
        \end{bmatrix}. \\
\end{split}
\end{equation} 

\bibliographystyle{elsarticle-num}
\bibliography{References}

\end{document}